\documentstyle[12pt]{article}

\skewchar\fivmi='177
\skewchar\sixmi='177
\skewchar\sevmi='177
\skewchar\egtmi='177
\skewchar\ninmi='177
\skewchar\tenmi='177
\skewchar\elvmi='177
\skewchar\twlmi='177
\skewchar\frtnmi='177
\skewchar\svtnmi='177
\skewchar\twtymi='177
\def\@magscale#1{ scaled \magstep #1}
\font\twfvmi  = ammi10   \@magscale5 
\skewchar\twfvmi='177
\skewchar\fivsy='60
\skewchar\sixsy='60
\skewchar\sevsy='60
\skewchar\egtsy='60
\skewchar\ninsy='60
\skewchar\tensy='60
\skewchar\elvsy='60
\skewchar\twlsy='60
\skewchar\frtnsy='60
\skewchar\svtnsy='60
\skewchar\twtysy='60
\font\twfvsy  = amsy10   \@magscale5 
\skewchar\twfvsy='60

\catcode`@=11
\def\un#1{\relax\ifmmode\@@underline#1\else
        $\@@underline{\hbox{#1}}$\relax\fi}
\catcode`@=12

\def\a{\alpha}
\def\b{\beta}

\def\d{\delta}
\def\e{\epsilon}

\def\g{\gamma}

\def\k{\kappa}
\def\l{\lambda}

\font\sc=font005                        
\font\ooo=circle10                      
\font\ro=manfnt                         
\def\kcl{{\hbox{\ro 6}}}                
\def\kcr{{\hbox{\ro 7}}}                
\def\ktl{{\hbox{\ro \char'134}}}        
\def\ktr{{\hbox{\ro \char'135}}}        
\def\kbl{{\hbox{\ro \char'136}}}        
\def\kbr{{\hbox{\ro \char'137}}}        

\def\bo{{\raise.15ex\hbox{\large$\Box$}}}               
\def\TH{{\raise.2ex\hbox{$\displaystyle \bigodot$}\mskip-4.7mu \llap H \;}}
\def\face{{\raise.2ex\hbox{$\displaystyle \bigodot$}\mskip-2.2mu \llap {$\ddot
        \smile$}}}                                      

\def\sp#1{{}^{#1}}                              
\def\leftrightarrowfill{$\mathsurround=0pt \mathord\leftarrow \mkern-6mu
        \cleaders\hbox{$\mkern-2mu \mathord- \mkern-2mu$}\hfill
        \mkern-6mu \mathord\rightarrow$}
\def\dvec#1{\vbox{\ialign{##\crcr
        \leftrightarrowfill\crcr\noalign{\kern-1pt\nointerlineskip}
        $\hfil\displaystyle{#1}\hfil$\crcr}}}           

\def\frac#1#2{{\textstyle{#1\over\vphantom2\smash{\raise.20ex
        \hbox{$\scriptstyle{#2}$}}}}}                   
\def\sfrac#1#2{{\vphantom1\smash{\lower.5ex\hbox{\small$#1$}}\over
        \vphantom1\smash{\raise.4ex\hbox{\small$#2$}}}} 
\def\bfrac#1#2{{\vphantom1\smash{\lower.5ex\hbox{$#1$}}\over
        \vphantom1\smash{\raise.3ex\hbox{$#2$}}}}       
\def\afrac#1#2{{\vphantom1\smash{\lower.5ex\hbox{$#1$}}\over#2}}    

\newskip\humongous \humongous=0pt plus 1000pt minus 1000pt
\def\caja{\mathsurround=0pt}

\newif\ifdtup
\def\panorama{\global\dtuptrue \openup2\jot \caja
        \everycr{\noalign{\ifdtup \global\dtupfalse
        \vskip-\lineskiplimit \vskip\normallineskiplimit
        \else \penalty\interdisplaylinepenalty \fi}}}
\def\eqalignno#1{\panorama \tabskip=\humongous                  
        \halign to\displaywidth{\hfil$\displaystyle{##}$
        \tabskip=0pt&$\displaystyle{{}##}$\hfil
        \tabskip=\humongous&\llap{$##$}\tabskip=0pt
        \crcr#1\crcr}}

\def\ref#1{$\sp{#1)}$}

\parskip=\medskipamount                 
\thicklines                         

\def\oldheadpic{                                
        \setlength{\unitlength}{.4mm}
        \thinlines
        \par
        \begin{picture}(349,16)
        \put(325,16){\line(1,0){4}}
        \put(330,16){\line(1,0){4}}
        \put(340,16){\line(1,0){4}}
        \put(335,0){\line(1,0){4}}
        \put(340,0){\line(1,0){4}}
        \put(345,0){\line(1,0){4}}
        \put(329,0){\line(0,1){16}}
        \put(330,0){\line(0,1){16}}
        \put(339,0){\line(0,1){16}}
        \put(340,0){\line(0,1){16}}
        \put(344,0){\line(0,1){16}}
        \put(345,0){\line(0,1){16}}
        \put(329,16){\oval(8,32)[bl]}
        \put(330,16){\oval(8,32)[br]}
        \put(339,0){\oval(8,32)[tl]}
        \put(345,0){\oval(8,32)[tr]}
        \end{picture}
        \par
        \thicklines
        \vskip.2in}
\def\oldtitle#1#2#3#4{\oldheadpic\begin{center}\vglue.5in{\large\bf #1}\\[.6in]
        {#2}\\[.1in] {\it Department of Physics and Astronomy}\\
        {\it University of Maryland, College Park, MD 20742}\\[.6in]
        Physics Publication \#{#3}\\ {#4}\\[1.5in] {\bf ABSTRACT}\\[.1in]
        \end{center} \begin{quotation}}                 
\def\oldTitle#1#2#3#4#5#6#7{\oldheadpic\begin{center} \vglue .4in
        {\large\bf #1}\\[.4in]
        {#2}\\[.1in] {\it Department of Physics and Astronomy}\\
        {\it University of Maryland, College Park, MD 20742}\\[.1in]
        {#3}\\[.1in] {\it {#4}}\\ {\it {#5}}\\[.4in]
        Physics Publication \#{#6}\\ {#7}\\[.5in] {\bf ABSTRACT}\\[.1in]
        \end{center} \begin{quotation}}                 
\def\border{                                            
        \setlength{\unitlength}{1mm}
        \newcount\xco
        \newcount\yco
        \xco=-24
        \yco=12
        \begin{picture}(140,0)
        \put(\xco,\yco){$\ktl$}
        \advance\yco by-1
        {\loop
        \put(\xco,\yco){$\kcl$}
        \advance\yco by-2
        \ifnum\yco>-240
        \repeat
        \put(\xco,\yco){$\kbl$}}
        \xco=158
        \yco=12
        \put(\xco,\yco){$\ktr$}
        \advance\yco by-1
        {\loop
        \put(\xco,\yco){$\kcr$}
        \advance\yco by-2
        \ifnum\yco>-240
        \repeat
        \put(\xco,\yco){$\kbr$}}
        \put(-20,11){\tiny University of Maryland Elementary Particle
Physics University of Maryland Elementary Particle Physics University of
Maryland Elementary Particle Physics}
        \put(-20,-241.5){\tiny University of Maryland Elementary
Particle Physics University of Maryland Elementary Particle Physics
University of Maryland Elementary Particle Physics}
        \end{picture}
        \par\vskip-8mm}
\def\bordero{                                           
        \setlength{\unitlength}{1mm}
        \newcount\xco
        \newcount\yco
        \xco=-24
        \yco=12
        \begin{picture}(140,0)
        \put(\xco,\yco){$\ktl$}
        \advance\yco by-1
        {\loop
        \put(\xco,\yco){$\kcl$}
        \advance\yco by-2
        \ifnum\yco>-240
        \repeat
        \put(\xco,\yco){$\kbl$}}
        \xco=158
        \yco=12
        \put(\xco,\yco){$\ktr$}
        \advance\yco by-1
        {\loop
        \put(\xco,\yco){$\kcr$}
        \advance\yco by-2
        \ifnum\yco>-240
        \repeat
        \put(\xco,\yco){$\kbr$}}
        \put(-20,12){\ooo
bacdefghidfghghdhededbihdgdfdfhhdheidhdhebaaahjhhdahba

hgdedge
   hgfdiehhgdigicba}
        \put(-20,-241.5){\ooo
ababaighefdbfghgeahgdfgafagihdidihiidhiagfedhadbfd

ecdcdfa
   gdcbhaddhbgfchbgfdacfediacbabab}
        \end{picture}
        \par\vskip-8mm}
\def\headpic{                                           
        \indent
        \setlength{\unitlength}{.4mm}
        \thinlines
        \par
        \begin{picture}(29,16)
        \put(165,16){\line(1,0){4}}
        \put(170,16){\line(1,0){4}}
        \put(180,16){\line(1,0){4}}
        \put(175,0){\line(1,0){4}}
        \put(180,0){\line(1,0){4}}
        \put(185,0){\line(1,0){4}}
        \put(169,0){\line(0,1){16}}
        \put(170,0){\line(0,1){16}}
        \put(179,0){\line(0,1){16}}
        \put(180,0){\line(0,1){16}}
        \put(184,0){\line(0,1){16}}
        \put(185,0){\line(0,1){16}}
        \put(169,16){\oval(8,32)[bl]}
        \put(170,16){\oval(8,32)[br]}
        \put(179,0){\oval(8,32)[tl]}
        \put(185,0){\oval(8,32)[tr]}
        \end{picture}
        \par\vskip-6.5mm
        \thicklines}
\def\title#1#2#3#4{\border\headpic {\hbox to\hsize{#4 \hfill UMDEPP #3}}\par
        \begin{center} \vglue .5in {\large\bf #1}\\[.6in]
        {#2}\\[.1in] {\it Department of Physics and Astronomy}\\
        {\it University of Maryland, College Park, MD 20742}\\[1.5in]
        {\bf ABSTRACT}\\[.1in] \end{center} \begin{quotation}}  
\def\Title#1#2#3#4#5#6#7{\border\headpic
        {\hbox to\hsize{#7 \hfill UMDEPP #6}}\par
        \begin{center} \vglue .4in {\large\bf #1}\\[.4in]
        {#2}\\[.1in] {\it Department of Physics and Astronomy}\\
        {\it University of Maryland, College Park, MD 20742}\\[.1in]
        {#3}\\[.1in] {\it {#4}}\\ {\it {#5}}\\[.5in] {\bf ABSTRACT}\\[.1in]
        \end{center} \begin{quotation}}                 
\def\endtitle{\end{quotation}\newpage}                  

\def\sect#1{\bigskip\medskip \goodbreak \noindent{\bf {#1}} \nobreak \medskip}
\def\refs{\sect{REFERENCES} \footnotesize \frenchspacing \parskip=0pt}
\def\Item{\par\hang\textindent}

\begin{document}


\def\gfrac#1#2{\frac {\scriptstyle{#1}}
        {\mbox{\raisebox{-.6ex}{$\scriptstyle{#2}$}}}}
\def\gg{{\hbox{\sc g}}}
\border\headpic {\hbox to\hsize{
August 1991 \hfill {UMDEPP 92-035}}}\par
\begin{center}
\vglue .4in
{\large\bf On the Perturbations of String--Theoretic Black Holes}
\\[.6in]
Gerald Gilbert\footnote {Research supported in part by NSF grant
\# PHY--91-41926}
\footnote {Bitnet: gng@umdhep ; Internet: gng@umdhep.umd.edu}\\[.3in]
{\it Department of Physics and Astronomy\\
University of Maryland at College Park\\
College Park, MD 20742-4111 USA}\\[.15in]

{\bf ABSTRACT}\\[.3in]
\end{center}
\begin{quotation}

The perturbations of string--theoretic black holes are analyzed by
generalizing the method of Chandrasekhar. Attention is focussed on
the case of the recently considered charged string--theoretic black
hole solutions as
a representative example. It is shown that string--intrinsic effects
greatly alter the perturbed motions of the string--theoretic black
holes as compared to the perturbed motions of black hole
solutions of the field equations of general relativity, the
consequences of which bear on the questions of the scattering behavior
and the stability of
string--theoretic black holes. The explicit forms of the axial potential
barriers surrounding the string--theoretic black hole are derived.
It is demonstrated that one of these, for sufficiently negative values
of the asymptotic value of the dilaton field, will inevitably
become negative in
turn, in marked contrast to the potentials surrounding the
static black holes of general relativity.
Such potentials may in principle be used in some cases to
obtain approximate
constraints on the value of the string coupling constant.
The application of the perturbation analysis to the case of
two--dimensional string--theoretic black holes is discussed.

\endtitle

\sect{{\bf 1. INTRODUCTION}}

\noindent{Recent research has disclosed the exciting
possibility of finding both exact and approximate
black hole solutions to
the still--unwritten equations
of motion in string theory [1,2,3] (see also [4,5]).
The
construction of exact conformal field theories yielding string
solutions in diverse dimensions corresponding to field configurations
surrounded by event horizons, as well
as other constructions, is
proceeding apace [6,7,8,9,10,11,12]. In addition
to aiding in the ongoing search for the Grail of
Nonperturbative Understanding in
string theory, the study of such configurations may also provide for
a deeper comprehension of the character of the solutions to the field
equations of the general theory of relativity which, after three--quarters
of a century of study, are still capable of harboring unexpected
surprises [13].}

\noindent{While the uncovering of new solutions in string theory,
with or without
event horizons, is an important enterprise, it is clearly of great
importance as well to investigate certain basic and requisite
features of such
discoveries. In this manner one hopes to be able to distinguish
between physically
robust solutions, and those mathematical solutions which
are in fact afflicted with characteristics which make them unsuitable
physically. In the case of black holes in general relativity,
for example, numerical and analytical investigations of
perturbations in the fields in the interior and the exterior
vicinity of particular solutions have revealed that certain putative
black holes are in fact unstable [14,15,16]. Other issues of black
hole persistence have in turn been raised by such considerations
[17,18,19,20].}

\noindent{The string--theoretic black hole solutions are, of
course, marked by features which should distinguish them from
the known solutions in general relativity. For instance, black holes
which are dyons may appear in general relativity as variants of the
original Reissner--Nordstr\"om solution, but the black hole dyons
which have thus far been discovered as (approximate) solutions to
string theory are composed of additional building--blocks comprised
by the axion or by the complex axidilaton [6,9]. In view of this we may
enquire as to what
departures will arise from that dynamical behavior expected
on the basis of
our experience with general relativity. To this end it
is crucial to carry out a careful analysis of the perturbed field
configurations associated with a given solution, taking care to
allow for the most general possible perturbed motions consistent with
all constraints, insofar as they are known, as dictated by string theory.}

\noindent{It is this problem to which we address our attention
in this paper, concerning
both four--dimensional and two--dimensional string--theoretic black holes.
In analyzing the perturbations of different solutions it will be
important to consider configurations determined in reference to the
metric tensor which is specified by
the underlying, defining
sigma--model. However, it is also worthwhile to utilize the Einstein
metric for such an analysis, particularly for purposes of comparison with
the solutions of general relativity.}

\noindent{While studies of the stability to various perturbations of
black holes have been carried out in different ways by various authors,
beginning with Regge and Wheeler [21], and
continuing thereafter by others [22,23,24,25],
we believe that the clearest and most coherent treatment has been
worked through by, primarily, Chandrasekhar [26,27,28].
In this paper we shall
generalize the method of Chandrasekhar to the case of the black holes
which may arise in string theory, and we will
apply this comprehensively
to a particular charged string--theoretic solution
as a representative example. We will discover that
string--intrinsic effects can, for certain ranges of values of
characteristic
parameters such as the asymptotic strength of the dilaton field,
substantially alter the behavior of string--theoretic black holes
as compared to the black holes of general relativity.
We will derive
explicit expressions for the potentials which surround string--theoretic
black holes (eq.'s (67) through (71)) which may be utilized as precise
tools to study the dynamical behavior. Although the expressions are
rather complicated, we will see that there are significant and
fundamental features inherent in (one of) these potentials which sharply
distinguish the string--theoretic black holes from the static black
holes of general relativity. Such effects in turn may induce marked
changes in the scattering behavior and stability properties of black
holes in string theory.}

\noindent{The outline of the paper is as follows. In Section(2a) we
derive the equations for the perturbed field components of
the electrically--charged string--theoretic black hole; in
Section(2b) we explicitly derive the potentials for axial perturbations;
in Section(3) we consider the polar perturbations and the
magnetically--charged black hole; in Section(4) we begin the analysis
of the perturbations of various two--dimensional black holes; Section(5)
contains concluding remarks, including comments regarding the
implications of our results for the scattering behavior and the
stability of black hole solutions in string theory.}

\newpage

\sect{{\bf 2. PERTURBATIONS OF THE ELECTRICALLY--CHARGED BLACK HOLE}}

\noindent{[{\it Conventions}: We shall adhere to the
following conventions: for black holes in four dimensions
the designation of the coordinates shall be
$\left(x^0, x^1, x^2, x^3\right) = \left(t, \varphi, r, \theta\right)$
(and {\it not} $\left(t, r, \theta, \varphi\right)$), the metric tensor
will always have the signature $\left(-, +, \cdots, +\right)$, and the
Levi--Civita tensor density $\epsilon_{\mu\nu\rho\sigma}$
is defined through $\epsilon_{0123}
={\sqrt {-g}}~[0123]$, where $[0123] = [t~\varphi~r~\theta]\equiv +1$.
In the analysis of two--dimensional black holes we shall use the
assignment $\left(x^0, x^1\right) = \left(t, r\right)$ for the
coordinates.
Also, for four--dimensional black holes we define
the Ricci tensor as the contraction of the second and
last indices of the
Riemann tensor, and in the case of two--dimensional black holes
we contract on the first and last indices.]}

\sect{{\bf 2a. EQUATIONS FOR THE PERTURBED FIELD COMPONENTS}}

\noindent{We begin by considering the string--inspired effective action}

$$S=\int d^4x {\sqrt {-g}}\left[-R+2\left(\nabla \Phi\right)^2+e^{-2\Phi}
F^2\right] ~,\eqno(1)$$

\noindent{where $F_{\mu\nu}$ is the Maxwell tensor and $\Phi=\Phi(r)$
is the
dilaton field. In this analysis we will not consider the effects of other
gauge fields or the antisymmetric tensor field. The equations of motion
which may be derived from the above action upon extremization with respect
to the various fields are}

$$\nabla_\mu\left(e^{-2\Phi}F^{\mu\nu}\right) = 0 ~,\eqno(2)$$

$$\nabla^2\Phi + {1\over 2}e^{-2\Phi}F^2 = 0 ~,\eqno(3)$$

$$R_{\mu\nu} = 2\nabla_\mu\Phi\nabla_\nu\Phi+2e^{-2\Phi}F_{\mu\rho}
F_{\nu}^{~\rho}-{1\over 2}e^{-2\Phi}g_{\mu\nu}F^2 ~,\eqno(4)$$

\noindent{while the other half of the Maxwell equations associated with eq.(2)
are}

$$\nabla_{\mu}\left(^*F^{\mu\nu}\right) = 0 ~,\eqno(5)$$

\noindent{which follow automatically from the definition of $F_{\mu\nu}=
\partial_\mu A_\nu - \partial_\nu A_\mu$,
in which the dual form $^*F^{\mu\nu}\equiv {1\over 2!}\epsilon^{\mu\nu\rho
\sigma}F_{\rho\sigma}$ defines the ordinary ($\Phi$--independent)
$^*$--duality operation.
In [5,10] a solution of the coupled equations (2) through (4)
is sought which is consistent with the requirements of static spherical
symmetry and the assumption of a purely magnetic field corresponding to
a Maxwell form given by $F=Q\sin\theta~d\theta\wedge d\varphi$,where Q is
the charge of the gauge field. The resulting black hole solution
to these equations is given by the following configuration of the fields:}

$$ds^2=-e^{2f_0}dt^2+e^{2f_1}d\varphi^2+e^{2f_2}dr^2+e^{2f_3}d\theta^2
{}~,\eqno(6)$$

$$e^{-2\Phi}=e^{-2\Phi_0}-{Q^2\over Mr} ~,\eqno(7)$$

\noindent{and}

$$F=Q\sin\theta~d\theta\wedge d\varphi ~\eqno(8)$$

\noindent{where $\Phi_0$ is the asymptotic (and also constant) value of
the dilaton field, and}

$$e^{2f_0}=e^{-2f_2}=1-{2M\over r} ~,\eqno(9)$$

$$e^{2f_3}=r\left(r-{Q^2e^{2\Phi_0}\over M}\right) ~,\eqno(10)$$

\noindent{and}

$$e^{2f_1}=e^{2f_3}\sin^2\theta ~.\eqno(11)$$

\noindent{While the above configuration is a static, spherically--symmetric
black hole solution characterized by a purely magnetic charge, it is
possible to derive a corresponding static, spherically--symmetric black
hole with a purely electric charge through the device of the
$\Phi$--dependent tilde--duality operation}

$${\tilde F}^{\mu\nu}\equiv {1\over 2!}e^{-2\Phi}\epsilon^{\mu\nu\rho\sigma}
F_{\rho\sigma} ~.\eqno(12)$$

\noindent{Now, by making the replacements $F_{\mu\nu}
\rightarrow {\tilde F}_{\mu\nu}$, $\Phi\rightarrow -\Phi$ and retaining
unchanged the metric tensor, the equations of motion (supplemented by
eq.(5)) transform into themselves (that this may occur in general
is contingent upon the absence of
source currents for the electromagnetic field), allowing for an
electrically--charged black hole solution. (This in spite of the fact
that, as may be readily verified,
the {\it action} is not invariant under the above
field transformation.) Performing the transformation, we obtain
({\it n.b.} the minus sign in eq.(3$'$)):}

$$\nabla_\mu\left(e^{+2\Phi}{\tilde F}^{\mu\nu}\right)=0 ~,\eqno(2')$$

$$\nabla^2\Phi-{1\over 2}e^{+2\Phi}{\tilde F}^2=0 ~,\eqno(3')$$

$$R_{\mu\nu}=2\nabla_\mu\Phi\nabla_\nu\Phi+2e^{+2\Phi}{\tilde F}_{\mu
\rho}{\tilde F}_\nu^{~\rho}-{1\over 2}e^{+2\Phi}g_{\mu\nu}{\tilde F}^2
{}~,\eqno(4')$$

\noindent{and}

$$\nabla_\mu\left(^*{\tilde F}^{\mu\nu}\right)=0 ~;\eqno(5')$$

\noindent{(which indeed reduce to eq.'s (2) through (5) upon
re--expressing $\tilde F$ in terms of $F$).}

\noindent{We will consider first the perturbed motions of the
electrically--charged black hole. For this purpose it will be
useful and convenient to introduce the antisymmetric
2--form $\cal F$:}

$${\tilde F}_{\mu\nu}\equiv -e^{-2\Phi}{\cal F}_{\mu\nu} ~,\eqno(13)$$

\noindent{where $\cal F$ is naturally defined as a curl through
${\cal F}_{\mu\nu}=
\partial_\mu{\cal A}_\nu-\partial_\nu{\cal A}_\mu$ with ${\cal A}_0=
Q\int e^{-2f_3} dr$, so that the only independent non--vanishing
component of ${\cal F}_{\mu\nu}$ is}

$${\cal F}_{02}=-Qe^{-2f_3} ~.\eqno(14)$$

\noindent{In terms of the field--strength ${\cal F}_{\mu\nu}$ eq.'s
(2$'$) through (5$'$) become}

$$\nabla_\mu{\cal F}^{\mu\nu}=0 ~,\eqno(2'')$$

$$\nabla_\mu\left(e^{-2\Phi}~{^*{\cal F}}^{\mu\nu}\right)=0 ~,\eqno(5'')$$

$$\nabla^2\Phi-{1\over 2}e^{-2\Phi}{\cal F}^2=0 ~,\eqno(3'')$$

$$R_{\mu\nu}=2\nabla_\mu\Phi\nabla_\nu\Phi+2e^{-2\Phi}{\cal F}_{\mu
\rho}{\cal F}_\nu^{~\rho}-{1\over 2}e^{-2\Phi}g_{\mu\nu}{\cal F}^2
{}~,\eqno(4'')$$

\noindent{The above four equations form the starting point for the
analysis of the perturbations of the electrically--charged
string--theoretic black hole.}

\noindent{To proceed we must address ourselves to the question of what
is the most
convenient and sufficiently general form of the perturbed metric tensor,
and what are the most sufficiently general perturbations of the other fields,
for our problem. As we are dealing with a static, spherically--symmetric
metric tensor, the perturbed configuration will be stationary (i.e.,
time--dependent) and without
spherical symmetry. However, it is clear that the most general
first--order perturbations
of an initially spherically--symmetric metric tensor will retain
a certain symmetry about some axis: it will suffice
throughout the analysis to restrict our
attention to {\it axially--symmetric} perturbations of the metric
tensor and other fields.{\footnote{In this connection we may observe
that the string--theoretic dyonic black hole considered in [6] was
not properly analyzed as to stability, as only spherically-symmetric
perturbations were considered by the authors. The method of
the present paper must be applied in order to complete the analysis
and unambiguously establish (or disprove) stability.}}
As will become evident in the sequel, this
fortunate fact results in the level of algebraic complexity of the
problem being reduced from one of likely
intractability to one of merely stupendous
intricacy.{\footnote{Although all calculations reported here were done by
hand, the computer program {\it Mathematica} [29] was found to be
invaluable as a
means of checking results.}} That metric of sufficient generality consistent
with time--dependent, axially--symmetric perturbations of the form of metric
given in eq.(6) has
been considered by Bardeen [25], and by
Chandrasekhar and Friedman [26], whose analysis yields:}

$$g^{\mu\nu}=\pmatrix{-e^{-2f_0}&-\chi_0e^{-2f_0}&0&0\cr
-\chi_0e^{-2f_0}&g^{11}&\chi_2e^{-2f_2}&\chi_3e^{-2f_3}\cr
0&\chi_2e^{-2f_2}&e^{-2f_2}&0\cr
0&\chi_3e^{-2f_3}&0&e^{-2f_3}\cr} ~,\eqno(15)$$

\noindent{for the contravariant form,}

$$g_{\mu\nu}=\pmatrix{\left(e^{2f_1}\chi_0^2-e^{2f_0}\right)&
-e^{2f_1}\chi_0&e^{2f_1}\chi_0\chi_2&e^{2f_1}\chi_0\chi_3\cr
-e^{2f_1}\chi_0&e^{2f_1}&-e^{2f_1}\chi_2&-e^{2f_1}\chi_3\cr
e^{2f_1}\chi_0\chi_2&-e^{2f_1}\chi_2&\left(e^{2f_1}\chi_2^2+
e^{2f_2}\right)&e^{2f_1}\chi_2\chi_3\cr
e^{2f_1}\chi_0\chi_3&-e^{2f_1}\chi_3&e^{2f_1}\chi_2\chi_3&
\left(e^{2f_1}\chi_3^2+e^{2f_3}\right)\cr} ~,\eqno(16)$$

\noindent{for the covariant form, where $g^{11}=\chi_3^{~2}e^{-2f_3}+
\chi_2^{~2}e^{-2f_2}+e^{-2f_1}-\chi_0^{~2}e^{-2f_0}$, and $\chi_0$, $\chi_2$
and $\chi_3$ are arbitrary functions (all assumed to be of the first--order
of smallness)
of $t$, $r$ and $\theta$, corresponding to a squared line
element given by}

$$ds^2=-e^{2f_0}dt^2+e^{2f_1}\left(d\varphi-\chi_0dt-\chi_2dr-
\chi_3d\theta\right)^2+e^{2f_2}dr^2+e^{2f_3}d\theta^2 ~.\eqno(17)$$

\noindent{Some algebra establishes the useful fact
that for the perturbed metric
tensor of eq.(16) one has}

$${\sqrt {-g}}=e^{f_0+f_1+f_2+f_3} ~,\eqno(18)$$

\noindent{which is the same result one obtains for the unperturbed
metric tensor given in eq.(6). Taking note of the explicit expression in
eq.(16) we observe that the general perturbation of the set of fields
given by eq.(6), (7) and}

$${\cal F}=-Qe^{-2f_3}~dt\wedge dr ~,\eqno(19)$$

\noindent{consists in: for the dilaton field
letting $\Phi\rightarrow \Phi+\delta\Phi$; for the electromagnetic field
letting
${\cal F}_{02}\rightarrow {\cal F}_{02}+\delta{\cal F}_{02}$, and $0
\rightarrow \delta{\cal F}_{\mu\nu}$ for the initially--vanishing remaining
independent components of the Maxwell tensor; and for
the gravitational field letting
$f_0\rightarrow f_0+\delta f_0, f_1\rightarrow f_1+\delta f_1, f_2\rightarrow
f_2+\delta f_2, f_3\rightarrow f_3+\delta f_3$, and $0\rightarrow \chi_0,
0\rightarrow \chi_2, 0\rightarrow \chi_3$. All quantities preceeded by a
$\delta$, as well as $\chi_0$, $\chi_2$ and $\chi_3$ are assumed to be of
the first order of smallness, and all of the perturbations depend in
general on $t$, $r$ and $\theta$.}

\noindent{The Chandrasekhar form{\footnote {Although the metric tensor
of eq.(16) was specifically considered in [25], the careful application of
this form to the analysis of the perturbations of particular solutions of the
field equations of general relativity was first
carried out in [26,27,28],
which motivates our choice of sobriquet.}}
of the perturbed metric (eq.(16)) is
cunningly chosen to exhibit an important fact. The general departures
from the static background may be put into two
classes ({\it cf.} eq.(17)):
those perturbations which upon a reversal of sign
leave the sign of the metric unchanged ($\delta f_0, \delta f_1,
\delta f_2$ and $\delta f_3$); and those perturbations which upon a reversal
of sign change the metric unless we let $\varphi\rightarrow
-\varphi$ as well ($\chi_0, \chi_2$ and $\chi_3$). Following Chandrasekhar
we shall refer to the two types of perturbations as {\it polar} for
the former and {\it axial} for the latter.{\footnote {The occurrence of
an axial perturbation generates the well--known dragging of the inertial
frame.}} The analysis to follow will
demonstrate that, as is the case for black holes in general relativity,
for electrically--charged
string--theoretic black holes the equations which determine the
two types of perturbations are entirely distinct and must {\it a priori}
be considered separately. An important consequence of this is
that, in general,
stability with respect to polar perturbations does not guarantee stability
with respect to axial perturbations, and {\it vice versa}.}

\noindent{Consistent with our assumption that the perturbations are
of first--order, we proceed to write down the coupled
Einstein--Maxwell--dilaton equations
for the fields, linearizing in the perturbations about the
static configuration. Turning our attention first to the Maxwell equations
(eq.'s (2$''$) and (5$''$)), we find (with the help of eq.'s(15) and (16))
after a moderate amount of algebra the following equations:}

$$\left(e^{f_1+f_2}\d{\cal F}_{12}\right)_{,3}+\left(e^{f_1+f_3}\d{\cal F}_{
31}\right)_{,2}=+2\left(e^{f_1+f_3}\d{\cal F}_{31}\Phi_{,2}+e^{f_1+f_2}
\d{\cal F}_{12}\Phi_{,3}\right) ~,\eqno(20a)$$

$$\left(e^{f_1+f_0}\d{\cal F}_{01}\right)_{,2}+\left(e^{f_1+f_2}\d{\cal F}_{
12}\right)_{,0}=+2\left(e^{f_1+f_0}\d{\cal F}_{01}\Phi_{,2}+e^{f_1+f_2}
\d{\cal F}_{12}\Phi_{,0}\right) ~,\eqno(20b)$$

$$\left(e^{f_1+f_0}\d{\cal F}_{01}\right)_{,3}+\left(e^{f_1+f_3}\d{\cal F}_{
13}\right)_{,0}=+2\left(e^{f_1+f_0}\d{\cal F}_{01}\Phi_{,3}+e^{f_1+f_3}
\d{\cal F}_{13}\Phi_{,0}\right) ~,\eqno(20c)$$

$$\eqalignno{\left(e^{f_2+
f_3}\d{\cal F}_{01}\right)_{,0}&+\left(e^{f_0+f_3}\d{\cal F}_{
12}\right)_{,2}+\left(e^{f_0+f_2}\d{\cal F}_{13}\right)_{,3}\cr
&=e^{f_1+f_3}
{\cal F}_{02}\Xi_{02}+e^{f_1+f_2}\d{\cal F}_{03}\Xi_{03}-e^{f_1+f_0}
\d{\cal F}_{23}\Xi_{23} ~,&(20d)\cr}$$

\noindent{and}

$$\left(e^{f_1+f_3}{\cal F}_{02}\right)_{,2}+\left(e^{f_1+f_2}\d{\cal F}_{
03}\right)_{,3}=0 ~,\eqno(21a)$$

$$-\left(e^{f_1+f_0}\d{\cal F}_{23}\right)_{,2}+\left(e^{f_1+f_2}\d{\cal F}_{
03}\right)_{,0}=0 ~,\eqno(21b)$$

$$\left(e^{f_1+f_0}\d{\cal F}_{23}\right)_{,3}+\left(e^{f_1+f_3}{\cal F}_{
02}\right)_{,0}=0 ~,\eqno(21c)$$

$$\eqalignno{\left(e^{f_0+f_2}{\cal F}_{02}\right)_{,3}&-\left(e^{f_0+
f_3}\d{\cal F}_{03}\right)_{,2}+\left(e^{f_2+f_3}\d{\cal F}_{23}
\right)_{,0}\cr
&=e^{f_1+f_0}\d{\cal F}_{01}\Xi_{23}+e^{f_1+f_2}\d{\cal F}_{12}\Xi_{03}-
e^{f_1+f_3}\d{\cal F}_{13}\Xi_{02}\cr &+2\left(e^{f_0+f_2}{\cal F}_{02}
\Phi_{,3}-e^{f_0+f_3}\d{\cal F}_{03}\Phi_{,2}+e^{f_2+f_3}\d{\cal F}_{23}
\Phi_{,0}\right) ~,&(21d)\cr}$$

\noindent{where $\Xi_{23}=-\Xi_{32}\equiv \chi_{2,3}-\chi_{3,2}$,
$\Xi_{20}\equiv \chi_{2,0}-\chi_{0,2}$,
and $\Xi_{30}\equiv \chi_{3,0}-\chi_{0,3}$. (It may be helpful
to remark that eq.'s
(20$d$), (21$a$), (21$b$) and (21$c$) are obtained from eq.(2$''$),
while eq.'s (20$a$), (20$b$), (20$c$) and (21$d$) are derived from
eq.(5$''$).)
These expressions have been
grouped into sets of equations which are invariant to the
substitution $\varphi
\rightarrow -\varphi$ (eq.'s (21)) and equations which change sign under this
transformation
(eq.'s (20)).{\footnote {We observe the very non--symmetric appearance
of the derivatives of $\Phi$ in the two sets of equations (20)
and (21). This, along
with related asymmetries between the equations for polar and axial
perturbations which we will discover in the sequel, will result in
significant consequences for the stability and scattering behavior
of electrically-- (or magnetically--) charged
string--theoretic black holes which will sharply
distinguish them from the black holes of general relativity.}}
Effecting the linearization for both sets of
equations, we find that the perturbations are determined by:}

$$\left(e^{-f_0+f_3}\d{\cal F}_{12}\sin\theta\right)_{,\theta}+
\left(e^{2f_3}\d{\cal F}_{31}\sin\theta\right)_{,r}=+2e^{2f_3}\d{\cal F}_{
31}\Phi_{,r}\sin\theta ~,\eqno(20a')$$

$$\left(e^{f_0+f_3}\d{\cal F}_{01}\sin\theta\right)_{,r}+e^{-f_0+
f_3}\d{\cal F}_{12,0}\sin\theta=+2e^{f_0+f_3}\d{\cal F}_{01}\Phi_{,r}
\sin\theta ~,\eqno(20b')$$

$$\left(e^{f_0+f_3}\d{\cal F}_{01}\sin\theta\right)_{,\theta}+e^{2f_3}
\d{\cal F}_{13,0}\sin\theta=0 ~,\eqno(20c')$$

$$e^{-f_0+f_3}\d{\cal F}_{01,0}+\left(e^{f_0+f_3}\d{\cal F}_{12}\right)_{,r}
+\d{\cal F}_{13,\theta}=e^{2f_3}{\cal F}_{02}\Xi_{02}\sin\theta ~,
\eqno(20d')$$

\noindent{and}

$$e^{-f_0+f_3}\d{\cal F}_{03,0}=\left(e^{f_0+f_3}\d{\cal F}_{23}
\right)_{,r} ~,\eqno(21b')$$

$$\d{\cal F}_{02,0}-Qe^{-2f_3}\left(\d f_1+\d f_3\right)_{,0}+
e^{f_0-f_3}\left(\d{\cal F}_{23}\sin\theta\right)_{,\theta}\csc
\theta=0 ~,\eqno(21c')$$

$$\eqalignno{[\d{\cal F}_{02}-&Qe^{-2f_3}\left(\d f_0+\d f_2
\right)]_{,\theta}+\left(e^{f_0+f_3}\d{\cal F}_{30}\right)_{
,r}+e^{-f_0+f_3}\d{\cal F}_{23,0}\cr &=+2\left({\cal F}_{
02}\d\Phi_{,\theta}-e^{f_0+f_3}\d{\cal F}_{03}\Phi_{,r}\right)
 ~,&(21d')\cr}$$

\noindent{where we have used: that ${\cal F}_{02}$ is the only
component of the Maxwell tensor which doesn't vanish in the background,
that $\Phi$ depends only on the radial coordinate, that we must
independently vary $f_0$, $f_1$, $f_2$ and $f_3$ before using eq.'s (9)
and (11),
and we note that $\Xi$ is a quantity of first--order, and we therefore
have ignored terms which include factors such as $\d{\cal F}_{03}\Xi_{03}$.
We have also neglected to linearize eq.(21$a$), taking account of the fact
that this equation merely serves as the integrability condition for
eq.'s (21$b$) and (21$c$).{\footnote {When making reference to the
``linearized Maxwell equations" we are not being redundant, but rather,
emphasizing that we are working to first order in the {\it perturbed}
field components.}}}

\noindent{Turning now to the dilaton equation (eq.(3$''$)) we find after
a considerable amount of algebra the following single linearized
equation for the perturbations:}

$$\eqalignno{&e^{-2f_0}\delta\Phi_{,0,0}-e^{-2f_2}\left[\delta\Phi_{,r,r}
+\left(f_1+f_0-f_2+f_3\right)_{,r}\delta\Phi_{,r}\right]\cr &-e^{-2f_3}
\left[\delta\Phi_{,\theta,\theta}+\left(f_1+f_0+f_2
-f_3\right)_{,\theta}\delta\Phi_{,\theta}
\right]+2e^{-2\Phi}e^{-2f_0-2f_2}{\cal F}_{02}^{~2}\delta\Phi\cr
&+e^{-2f_2}\left\{\left(\delta f_1+\delta f_0-\delta f_2+\delta f_3
\right)_{,r}\Phi_{,r}-2\delta f_2\left[\Phi_{,r,r}+\left(f_1+f_0-f_2+
f_3\right)_{,r}\Phi_{,r}\right]\right\}\cr &-2e^{-2\Phi}e^{-2f_0-2f_2}
\left[{\cal F}_{02}\delta{\cal F}_{02}-\left(\delta f_0+\delta f_2\right)
{\cal F}_{02}^{~2}\right]=0 ~.&(22)\cr}$$

\noindent{Inspection of the above equation reveals that only the polar
perturbations appear. In the non--appearance of the axial perturbations
we see yet
another indication of what was suggested in footnote \#7 above:
the presence of the dilaton field upsets a balance which, in the case of
the static black holes of general relativity, relates the polar and axial
perturbations through the associated scattering amplitudes.}

\noindent{Turning next to the Einstein equations (eq.(4$''$)), we
must set the perturbations in the Ricci tensor equal to the
perturbations in the stress--energy tensor.
The stress-energy tensor for the problem is given by}

$$T_{\mu\nu}^{({\rm tot})}=T_{\mu\nu}^{(\Phi)}+e^{-2\Phi}T_{\mu\nu}^{
({\rm EM})} ~,\eqno(23)$$

\noindent{where}

$$T_{\mu\nu}^{(\Phi)}=2\nabla_\mu\Phi\nabla_\nu\Phi ~,\eqno(24)$$

\noindent{and}

$$T_{\mu\nu}^{({\rm EM})}=+2{\cal F}_{\mu\rho}{\cal F}_\nu^{~\rho}-
{1\over 2}g_{\mu\nu}{\cal F}^2 .~\eqno(25)$$

\noindent{Since the variation of this must be set equal to the
variation of the Ricci tensor (which depends in turn upon the
variations of the various metric components) we have explicitly:}

$$\eqalignno{\d T_{ab}^{({\rm tot})}&=2\left(\d\Phi_{,a}\Phi_{,b}+
\Phi_{,a}\d\Phi_{,b}\right)+
e^{-2\Phi}\left(\d T_{ab}^{
({\rm EM})}-2T_{ab}^{({\rm EM})}\d\Phi\right)\cr
&=2\left(\d\Phi_{,a}\Phi_{,b}+\Phi_{,a}\d\Phi_{,b}\right)\cr
&~~~+2e^{-2\Phi}\left[\eta^{cd}\left(\d{\cal F}_{ac}{\cal F}_{bd}+
{\cal F}_{ac}\d{\cal F}_{bd}\right)+\eta_{ab}{\cal F}_{02}\d{\cal F}_{
02}-T_{ab}^{({\rm EM})}\d\Phi\right] ~,&(26)\cr}$$

\noindent{where the lower--case roman letters indicate that the
tensor components
are referred to an orthonormal vierbein frame $e_a^{~k}$
with Minkowski metric
tensor $e_a^{~k}e_{bk}=\eta_{ab}$
of signature $(-+++)$, consistent with the signature of the
metric tensor of eq.(6), so that contractions are implicitly performed
in this frame, and thus}

$$T_{ab}^{({\rm EM})}=2{\cal F}_{ac}{\cal F}_b^{~c}+\eta_{ab}Q^2e^{
-4f_3} ~.\eqno(27)$$

\noindent{It will be necessary in the subsequent analysis
to have available the explicit values for
certain components of $\d T_{ab}^{({\rm tot})}$ which we compute to be:}

$$\d T_{13}^{({\rm tot})}=0 ~,\eqno(28)$$

$$\d T_{12}^{({\rm tot})}=+2e^{-2\Phi}e^{-2f_3}Q\d{\cal F}_{01} ~,
\eqno(29)$$

$$\d T_{02}^{({\rm tot})}=+2\Phi_{,r}\d\Phi_{,0} ~,\eqno(30)$$

$$\d T_{03}^{({\rm tot})}=-2e^{-2\Phi}e^{-2f_3}Q\d{\cal F}_{32} ~,
\eqno(31)$$

$$\d T_{23}^{({\rm tot})}=+2\Phi_{,r}\d\Phi_{,\theta}+2e^{-2\Phi}
e^{-2f_3}Q\d{\cal F}_{03} ~,\eqno(32)$$

$$\d T_{11}^{({\rm tot})}=-2e^{-2\Phi}\left(Qe^{-2f_3}\d{\cal F}_{02}+
Q^2e^{-4f_3}\d\Phi\right) ~,\eqno(33)$$

$$\eqalignno{\d T_{22}^{({\rm tot})}&=4\d\Phi_{,2}\Phi_{,2}-2e^{
-2\Phi}\left(2\eta^{00}+\eta_{22}\right)\left(\d{\cal F}_{02}Q
e^{-2f_3}+\d\Phi Q^2e^{-4f_3}\right)\cr &=4\d\Phi_{,r}\Phi_{,r}
+2e^{-2\Phi}\left(Qe^{-2f_3}\d{\cal F}_{02}+Q^2e^{-4f_3}\d\Phi
\right) ~.&(34)\cr}$$

\noindent{Considering now the perturbations in the Ricci
tensor, it turns out that, as in the case of the Schwarzschild
black hole, the dynamical evolution of the axial perturbations $\chi_0$,
$\chi_2$ and $\chi_3$ is uniquely determined by the $(1,2)$-- and
$(1,3)$--components of the Einstein equations [26]. The corresponding
components of the vacuum Ricci tensor, referred to the
metric of eq.(16), may be worked out with some effort to be [26,25]:}

$$R_{12}={1\over 2}e^{-2f_1-f_0-f_3}\left[\left(e^{3f_1+f_0-f_2-f_3}
\Xi_{23}\right)_{,\theta}-\left(e^{3f_1-f_0+f_3-f_2}\Xi_{20}\right)_{
,0}\right] ~,\eqno(35)$$

\noindent{and}

$$R_{13}={1\over 2}e^{-2f_1-f_0-f_2}\left[\left(e^{3f_1+f_0-f_3-f_2}
\Xi_{32}\right)_{,r}-\left(e^{3f_1-f_0+f_2-f_3}\Xi_{30}\right)_{,0}
\right] ~.\eqno(36)$$

\noindent{Setting the corresponding components of the Ricci and
stress-energy tensors equal to each other and effecting the
linearization with the aid of eq.'s (28) and (29), we obtain:}

$$\left(e^{2f_0+2f_3}\Xi_{23}\sin^3\theta\right)_{,\theta}-
e^{4f_3}\Xi_{20,0}\sin^3\theta=4Qe^{-2\Phi}e^{f_0+f_3}\d{\cal F}_{
01}\sin^2\theta ~,\eqno(37)$$

\noindent{and}

$$\left(e^{2f_0+2f_3}\Xi_{23}\right)_{,r}-e^{-2f_0+2f_3}\Xi_{03,0}
=0 ~,\eqno(38)$$

\noindent{for the $(1,2)$-- and $(1,3)$--component equations,
respectively. Proceeding in the same manner for the polar
perturbations, we find we need to consider the $(2,2)$--, $(1,1)$--,
$(2,3)$--, $(0,3)$-- and $(0,2)$--components of the Einstein
equations [26]. Again equating the appropriate components of the
Ricci and stress--energy tensors, and linearizing in the
polar perturbations, we derive:}

$$\displaylines{\d R_{22}=\d T_{22}\Rightarrow\hfill\cr}$$
$$\eqalignno{&e^{2f_0}\left[2{f_3}_{,r}\d {f_0}_{,r}+\left({f_0}_{,r}+
{f_3}_{,r}\right)\left(\d f_1+\d f_3\right)_{,r}-2\left(2{f_0}_{,r}
{f_3}_{,r}+{f_3}_{,r}^{~2}\right)\d f_2\right]\cr &+e^{-2f_3}\left[
\left(\d f_1+\d f_0\right)_{,\theta ,\theta}+\left(\d f_0-\d f_3+
2\d f_1\right)_{,\theta}\cot\theta+2\d f_3\right]\cr &-e{-2f_0}
\left(\d f_1+\d f_3\right)_{,0,0}\cr &=4\Phi_{,r}\d\Phi_{,r}+
2e^{-2\Phi}\left(Qe^{-2f_3}\d{\cal F}_{02}+Q^2e^{-4f_3}\d\Phi
\right) ~,&(39)\cr}$$

$$\displaylines{\d R_{11}=\d T_{11}\Rightarrow\hfill\cr}$$
$$\eqalignno{&e^{2f_0}\{\d {f_1}_{,r,r}+{f_3}_{,r}
(\d f_1+\d f_0+\d f_3-\d f_2)_{,r}\cr &+2\d {f_1}_{,r}
({f_3}_{,r}+{f_1}_{,r})-2[{f_3}_{,r,r}+2{f_3}_{,r}
({f_3}_{,r}+{f_0}_{,r})]\d f_2\}\cr
&+e^{-2f_3}[\d {f_1}_{,\theta ,\theta}+(2\d f_1+\d f_0
+\d f_2-\d f_3)_{,\theta}\cot\theta +2\d f_3]\cr
&-e^{-2f_0}\d {f_1}_{,0,0}\cr &=2e^{-2\Phi}(Qe^{
-2f_3}\d{\cal F}_{02}+Q^2e^{-4f_3}\d\Phi) ~,&(40)\cr}$$

$$\displaylines{\d R_{23}=\d T_{23}\Rightarrow\hfill\cr}$$
$$\eqalignno{\left({f_0}_{,r}+{f_3}_{,r}\right)\d {f_2}_{,\theta}
&-\left(\d f_1+\d f_0\right)_{,r,\theta}-\left({f_0}_{,r}-{f_3}_{,r}
\right)\d {f_0}_{,\theta}-\left(\d f_1-\d f_3\right)_{,r}\cot\theta\cr
&=2e^{-f_0+f_3}\left(\Phi_{,r}\d\Phi_{,\theta}+e^{-2\Phi}e^{-2f_3}Q
\d{\cal F}_{03}\right) ~,&(41)\cr}$$

$$\displaylines{\d R_{03}=\d T_{03}\Rightarrow\hfill\cr}$$
$$\left(\d f_1+\d f_2\right)_{,\theta,0}+\left(\d f_1-\d f_3\right)_{,0}
\cot\theta=2e^{-2\Phi}e^{f_0-f_3}Q\d{\cal F}_{32} ~,\eqno(42)$$

$$\displaylines{\d R_{02}=\d T_{02}\Rightarrow\hfill\cr}$$
$$\left[\left(\d f_1+\d f_3\right)_{,r}+\left(\d f_1+\d f_3\right)\left(
{f_3}_{,r}-{f_0}_{,r}\right)-2{f_3}_{,r}\d f_2\right]_{,0}
=-2\Phi_{,r}\d\Phi_{,0} ~.\eqno(43)$$

\noindent{With these expressions for the perturbed Einstein equations
we may assemble all of the equations relevant to our analysis of
the perturbations of the charged string--theoretic black hole:}

\noindent{(1) The {\it axial} perturbations may be determined by
analyzing eq.'s (20$a'$) through (20$d'$) together with eq.'s (37)
and (38);}

\noindent{(2) The {\it polar} perturbations
may be determined by analyzing eq.'s (21$b'$) through (21$d'$) together
with eq.(22) and eq.'s (39) through (43).}

\noindent{We note finally that, just as only polar perturbations
appeared in the linearized dilaton equation, so we see that the
equations for the axial perturbations do not contain any terms
which include perturbations in the dilaton field. This lack of
symmetry is not surprising in view of the purely radial dependence
of the dilaton field.}

\sect{{\bf 2b. DERIVATION OF THE POTENTIAL FOR AXIAL PERTURBATIONS}}

\noindent{The axial perturbations are entirely determined by
the coupled partial
differential equations (20$a'$) through (20$d'$) along with
eq.'s (37) and (38). We begin
analyzing the axial potential by considering  the electromagnetic
equations.
In analyzing the Maxwell
equations it will suffice to consider eq.(20$b'$) through eq.(20$d'$).
Utilizing
eq.'s (20$b'$) and (20$c'$) to eliminate $\d{\cal F}_{12}$ and
$\d{\cal F}_{13}$
from eq.(20$d'$) we find (after a number of cancellations)}

$$\eqalignno{e^{-f_0+f_3}\b_{,0,0}&-
{\partial\over \partial r}\left[e^{2f_0}
{\partial\over \partial r}\left( e^{f_0+f_3}\b\right)\right]
+2{\partial\over \partial r}\left( e^{3f_0+f_3}\b\Phi_{,r}\right)\cr
&-e^{f_0-f_3}
{\partial\over \partial\theta}\left( {1\over \sin\theta}{\partial\b\over
\partial\theta}\right)
\sin\theta=e^{2f_3}{\cal F}_{02}\Xi_{02,0}\sin^2\theta ~,&(44)\cr}$$

\noindent{where we have introduced the abbreviation $\d{\cal F}_{01}\sin
\theta\equiv \b$. We may also combine the Einstein equations for $\delta
R_{12}$ and $\delta R_{13}$ to obtain the single partial differential
equation}

$$\eqalignno{e^{4f_3}{\partial\over \partial r}\left(e^{-2f_3+2f_0}
{\partial\a\over \partial r}\right)+\sin^3\theta{\partial\over \partial
\theta}\left({1\over \sin^3\theta}{\partial\a\over \partial\theta}\right)
&-\a_{,0,0}e^{2f_3-2f_0}\cr &=2e^{f_0+3f_3}\sin^3\theta{\partial\over
\partial\theta}\left({\delta T_{12}\over \sin\theta}\right) ~,&(45)\cr}$$

\noindent{where we have defined
$e^{2f_3+2f_0}\Xi_{23}\sin^3\theta\equiv \a$. In further
reducing the above coupled pair of partial differential equations we are
interested in the first instance in
separating the dependence on the coordinates in such a manner that the
separated functional forms for $\a$ and $\b$ are mutually consistent.
We will thereafter attempt to decouple the two resulting ordinary
differential equations. That this proves possible is
a consequence of our having chosen the electrically--charged black hole for
analysis. (We will see below that in the case of the magnetically--charged
black hole the simultaneous separation of variables and
subsequent decoupling appears not
to be possible, and we will be led to consider instead a pair of
coupled integrodifferential equations for the perturbations of the black
hole.) Employing the expression for $\delta T_{12}$ given in eq.(29)
in eq.(45), simultaneously separating the variables in eq.'s (44)
and (45) through the substitution}

$$\a(r,\theta)=\a(r)C_{l+2}^{-3/2}(\theta) ~,\eqno(46)$$

$$\b(r,\theta)=\b(r){dC_{l+2}^{-3/2}\over d\theta}\csc\theta
{}~,\eqno(47)$$

\noindent{where $C_n^m(\theta)$ is the Gegenbauer polynomial, and
performing a temporal Fourier analysis for a particular frequency
$\omega$ (henceforth we shall assume that all perturbations have
a time--dependence $\propto e^{i\omega t}$) we
may derive after some algebra the following coupled pair of
ordinary differential equations}

$$e^{2f_3+2f_0}{d\over dr}\left(e^{-2f_3+2f_0}{d\a\over dr}\right)
+\omega^2\a-e^{-2f_3+2f_0}n^2\a=-4n^2Qe^{-2\Phi}e^{3f_0-f_3}\b
{}~,\eqno(48)$$

$$\eqalignno{-2{d\over dr}\left(e^{3f_0+f_3}\b
\Phi_{,r}\right)&+{d\over dr}
\left[e^{2f_0}{d\over dr}\left(e^{f_0+f_3}\b\right)\right]
-\left(n^2+2\right)e^{f_0-f_3}\b\cr
&+\left(\omega^2e^{-f_0+f_3}-4Q^2e^{-2\Phi}
e^{f_0-3f_3}\right)\b=-e^{-4f_3}Q\a ~,&(49)\cr}$$

\noindent{where $n^2=(l-1)(l+2)$ is the separation constant, and
we have also made use of the following identity
which may be verified using eq.(37):}

$$\Xi_{02,0}\sin^2\theta=4Qe^{-2\Phi}e^{f_0-3f_3}\b-e^{-4f_3}
{\partial\a\over \partial\theta}\csc\theta ~.\eqno(50)$$

\noindent{To further reduce the coupled equations eq.(48) and eq.(49) we
recall the basic fact that for an ordinary differential equation
of the form}

$${d^2f\over dx^2}+X_1(x){df\over dx}+X_2(x)f=h$$

\noindent{we may eliminate the first--order term by employing the field
redefinition $f(x)\rightarrow A(x)g(x)$, where $g(x)$ is the new
dependent variable, and the integrating factor $A(x)$ is given by}

$$A(x)=\exp\left(-{1\over 2}\int^x X_1(x) dx\right) ~.\eqno(51)$$

\noindent{We also observe that, for the
charged string--theoretic black hole solution the
forms of the metric components $g_{00}$ and $g_{22}$ are
precisely the same as for
the Schwarzschild solution, and thus it is natural to introduce the
``luminosity distance" $r_*$ (first introduced by Regge and
Wheeler [21]) defined by}

$$r_*=r+2M\log({r\over 2M}-1) ~.\eqno(52)$$

\noindent{Substituting $r_*$ for $r$ in eq.'s (48) and (49) as the
independent variable and observing that}

$${d\over dr_*}=e^{2f_0}{d\over dr}$$

\noindent{we may deduce that the required integrating
factors $A$ are}

$$\eqalignno{A(r)&=\exp\left[-{1\over 2}\int {d\over dr_*}
\left(-2f_3\right)dr_*\right]\cr &=e^{f_3} ~,&(53)\cr}$$

\noindent{for eq.(48), and}

$$\eqalignno{A(r)&=\exp\left[-{1\over 2}\int 2{d\over dr_*}\left(
-\Phi+f_0+f_3\right)dr_*\right]\cr &=e^{\Phi}e^{-f_0-f_3} ~,&(54)\cr}$$

\noindent{for eq.(49). We are thus led to employ the field
redefinitions}

$$\a(r)=e^{f_3}Y_2(r) ~,\eqno(55)$$

\noindent{and}

$$\b(r)=-{1\over 2n}e^{\Phi}e^{-f_0-f_3}Y_1(r) ~,\eqno(56)$$

\noindent{in eq.'s (48) and (49). Carrying out the required
substitution results in the elimination of
the first--order terms and we arrive at the
following pair of coupled equations for the perturbations:}

$$\eqalignno{{\cal Q}^2Y_1=e^{2f_0-2f_3}&[(n^2+2-e^{-2f_0+
2f_3}{\cal T}_1+4Q^2e^{-2\Phi}e^{-2f_3})Y_1\cr &+2nQe^{-\Phi}
e^{-f_3}Y_2] ~,&(57)\cr}$$

\noindent{and}

$${\cal Q}^2Y_2=e^{2f_0-2f_3}\left[\left(n^2-e^{-2f_0+2f_3}{\cal T}_2
\right)Y_2+2nQe^{-\Phi}e^{-f_3}Y_1\right] ~,\eqno(58)$$

\noindent{in which we have found it useful for brevity to define:}

$${\cal T}_1\equiv {d^2\over dr_*^{~2}}\left(\Phi-f_0-f_3\right)-
\left[{d\over dr_*}\left(\Phi-f_0-f_3\right)\right]^2 ~,\eqno(59)$$

\noindent{and}

$${\cal T}_2\equiv {d^2\over dr_*^{~2}}f_3-\left({d\over dr_*}f_3
\right)^2 ~,\eqno(60)$$

\noindent{and we have also introduced the
Regge--Wheeler--Schr\"odinger operator [21]}

$${\cal Q}^2\equiv {d^2\over dr_*^{~2}}+\omega^2 ~.\eqno(61)$$

\noindent{We can neatly represent the pair of radial equations
eq.(57) and eq.(58) in the
following matrix form:}

$${\cal Q}^2\pmatrix{Y_1\cr Y_2\cr}=\k\pmatrix{A&C\cr C&B\cr}
\pmatrix{Y_1\cr Y_2\cr} ~,\eqno(62)$$

\noindent{where we have introduced}

$$\k\equiv e^{2f_0-2f_3} ~,\eqno(63)$$

$$A\equiv n^2+2-e^{-2f_0+2f_3}{\cal T}_1+4Q^2e^{-2\Phi}e^{-2f_3} ~,
\eqno(64)$$

$$B\equiv n^2-e^{-2f_0+2f_3}{\cal T}_2 ~,\eqno(65)$$

\noindent{and}

$$C\equiv 2nQe^{-\Phi}e^{-f_3} ~.\eqno(66)$$

\noindent{Thus we see that we may decouple the equations by solving
for the eigenvalues of this matrix equation (eq.(62)).
Performing the
diagonalization, we find after much algebra (and with the help
of eq.'s (7), (9) and (10) and after many reductions)
the following explicit expressions for the potentials,
associated with the axial perturbations, which surround the
electrically--charged string--theoretic black holes:}

$$V_1(r,Q,M,\Phi_0)=\l\left(v
+{\sqrt {\Delta}}\right) ~,\eqno(67)$$

\noindent{and}

$$V_2(r,Q,M,\Phi_0)=\l\left(v
-{\sqrt {\Delta}}\right) ~,\eqno(68)$$

\noindent{where:}

$$\l(r,Q,M,\Phi_0)=\left[-8e^{2\Phi_0}r^4\left(Mr-
Q^2e^{2\Phi_0}\right)^2\right]^{-1} ~,\eqno(69)$$

$$\eqalignno{v(r,Q,M,\Phi_0)=&-32M^2Q^4e^{2\Phi_0}
-40M^2Q^4e^{6\Phi_0}+(32M^3Q^2+16 MQ^4e^{2\Phi_0}\cr
&+64M^3Q^2e^{4\Phi_0}+24MQ^4e^{6\Phi_0})r-
(16M^2Q^2+36M^4e^{2\Phi_0}\cr
&+60M^2Q^2e^{4\Phi_0} +
3Q^4e^{6\Phi_0}+16M^2Q^2e^{4\Phi_0}n^2)r^2 +
(48M^3e^{2\Phi_0}\cr &+16MQ^2e^{4\Phi_0}
+16M^3e^{2\Phi_0}n^2+8MQ^2e^{4\Phi_0}n^2)r^3
-(16 M^2e^{2\Phi_0}\cr &+8M^2e^{2\Phi_0}n^2)r^4 ~,&(70)\cr}$$

\noindent{and}

$$\eqalignno{\Delta(r,Q,M,\Phi_0)=&1024M^4Q^8e^{4\Phi_0}\cr &+
(-2048M^5Q^6e^{2\Phi_0} -
1024M^3Q^8e^{4\Phi_0}+3072M^5Q^6e^{6\Phi_0}
+512M^3Q^8e^{8\Phi_0}) r\cr
&+(1024M^6 Q^4+2048M^4Q^6e^{2\Phi_0}-
6912M^6Q^4e^{4\Phi_0}+256M^2Q^8e^{4\Phi_0}\cr &-
3840M^4Q^6e^{6\Phi_0}+2304M^6Q^4e^{8\Phi_0}-
448M^2Q^8e^{8\Phi_0}+768 M^4 Q^6e^{10\Phi_0}\cr
&+64M^2Q^8e^{12\Phi_0}+1024M^4Q^6e^{6\Phi_0}n^2)r^2\cr &+
(-1024M^5Q^4+3840M^7Q^2e^{2\Phi_0}-512M^3Q^6e^{2\Phi_0}+
7296M^5Q^4e^{4\Phi_0}\cr &-5760M^7Q^2e^{6\Phi_0}+
1344M^3Q^6e^{6\Phi_0}-3648M^5Q^4e^{8\Phi_0}+96MQ^8e^{8\Phi_0}\cr &-
736M^3Q^6e^{10\Phi_0}-48MQ^8e^{12\Phi_0}-
2048M^5Q^4e^{4\Phi_0}n^2-1024M^3Q^6e^{6\Phi_0}n^2)r^3\cr
&+(256M^4Q^4-3968M^6Q^2e^{2\Phi_0}
+3600M^8e^{4\Phi_0}-1920M^4Q^4e^{4\Phi_0}\cr &+6432M^6Q^2e^{6\Phi_0}
-96 M^2Q^6e^{6\Phi_0}+
1656M^4Q^4e^{8\Phi_0}+ 168 M^2Q^6e^{10\Phi_0}\cr &+9Q^8e^{12\Phi_0}+
1024M^6Q^2e^{2\Phi_0}n^2+2048M^4Q^4e^{4\Phi_0}n^2+
256M^2Q^6e^{6\Phi_0}n^2)r^4\cr &+
(1024M^5Q^2e^{2\Phi_0}-3840M^7e^{4\Phi_0}-1792M^5Q^2e^{6\Phi_0}
-192 M^3Q^4e^{8\Phi_0}\cr &-1024M^5Q^2e^{2\Phi_0}n^2-
512 M^3Q^4e^{4\Phi_0}n^2)r^5\cr &+
(1024M^6e^{4\Phi_0}+256M^4Q^2e^{2\Phi_0}n^2)r^6 ~.&(71)\cr}$$

\noindent{We have thus reduced the analysis of the
axial perturbations of the string--theoretic black hole to
the consideration of the two Schr\"odinger equations:}

$${\cal Q}^2{\cal Z}_1=V_1{\cal Z}_1 ~,\eqno(72)$$

\noindent{and}

$${\cal Q}^2{\cal Z}_2=V_2{\cal Z}_2 ~,\eqno(73)$$

\noindent{where ${\cal Z}_1$ and ${\cal Z}_2$ are the eigenvectors
associated with the matrix of eq. (62).}

\noindent{Although a complete understanding of the implications
of the complicated expressions we have derived for the axial
potentials must await a thorough numerical analysis,
a preliminary investigation reveals
prominent departures from the expectations one has from
experience with the static black hole configurations of general
relativity. As pointed out in [10], the extremal value of the charge
of the gauge field for the string--theoretic black hole is not
the same as for the extremal value in the Reissner--Nordstr\"om
black hole: the numerical value in the former case is greater
(or lower, according to the sign of the asymptotic value
of the dilaton field) by a factor ${\sqrt {2}}e^{-\Phi_0}$.
The two axial
potentials one may derive for the Reissner--Nordstr\"om black hole
have the property that, as long as the charge is less than the
extremal value, and as long as one probes the region exterior
to the event horizon, they assume real, and in particular positive,
values
as functions of the radial coordinate.{\footnote {The (single) axial
potential surrounding the Schwarzschild black hole is also real and positive
in the region exterior to the event horizon.}}
The stability to external
perturbations of the
Reissner--Nordstr\"om black hole is a
consequence of this fact.
This property is shared
by the potential $V_1$ (eq.(67)) we have derived. However, one finds
that, even
for values of the charge {\it well within the extremal limit of
the Reissner--Nordstr\"om black hole}, the potential $V_2$ (eq.(68))
becomes
negative when the asymptotic value of the dilaton field becomes
sufficiently negative. When the charge is allowed to exceed the
extremal limit of the Reissner--Nordstr\"om black hole, but is constrained
to remain less than the extremal limit dictated by the string--theoretic
black hole (i.e., $Q^2<2e^{-2\Phi_0}M^2$) one finds that the potential
oscillates wildly between negative and positive values in the region
exterior to the event horizon. (More nodes develop as the
value of the dilaton field becomes more negative.)
For example, if the value of $M$ is
normalized to unity and the value of the charge
$Q$ is taken to be $0.4$ (well within the extremal Reissner-Nordstr\"om
limit), for the spherical harmonic of orbital quantum number
$l=2$ (recall
$n^2=(l-1)(l+2)$) the potential $V_2$ becomes negative for values
of $\Phi_0$ less than $\sim -6$. In Table 1 we have given a list of
typical values of the axial potential $V_2$ in the exterior vicinity
of the event horizon (corresponding to the value $r/M=2$)
of the charged black hole. With the exception
of the last column in which $Q=10^7$, which corresponds to a naked
singularity, all of the data are for $Q$-values within the extremal
limit dictated by string theory.}

\noindent{{\bf Table 1}: The Axial Potential
$V_2$: ($l=2$, $e^{2\Phi_0}=10^{-10}$, $\Phi_0\cong -11.5$)}

\medskip

\vbox{\offinterlineskip
\def\tablerule{\noalign{\hrule}}
\hrule
\halign{&\vrule#&
  \strut\quad\hfil#\quad\cr
height2pt&\omit&&\omit&&\omit&&\omit&&\omit&\cr
&$r/M$&&$Q=0.2$&&$Q=20$&&$Q=10^5$&&$Q=10^7$&\cr\tablerule
height2pt&\omit&&\omit&&\omit&&\omit&&\omit&\cr
&$2.0$&&$+3.125.10^{-2}$&&$0$&&$0$&&$0$&\cr\tablerule
height2pt&\omit&&\omit&&\omit&&\omit&&\omit&\cr
&$2.2$&&$-1.367.10^{-2}$&&$-1.388.10^{-2}$&&$+1.357.10^5$&&$-3.416.10^{18}$&\cr
height2pt&\omit&&\omit&&\omit&&\omit&&\omit&\cr\tablerule
height2pt&\omit&&\omit&&\omit&&\omit&&\omit&\cr
&$2.4$&&$-1.447.10^{-2}$&&$-1.482.10^{-2}$&&$+1.643.10^4$&&$-4.824.10^{18}$&\cr
\tablerule
&$2.6$&&$-1.050.10^{-2}$&&$-1.063.10^{-2}$&&$-1.435.10^4$&&$-5.253.10^{18}$&\cr
\tablerule
&$2.8$&&$-5.206.10^{-3}$&&$-5.322.10^{-3}$&&$-7.662.10^2$&&$-5.208.10^{18}$&\cr
\tablerule
&$3.0$&&$+6.853.10^{-9}$&&$-2.500.10^{-5}$&&$-1.884.10^3$&&$-4.940.10^{18}$&\cr
\tablerule
&$3.2$&&$+4.578.10^{-3}$&&$+4.540.10^{-3}$&&$-1.203.10^3$&&$-4.580.10^{18}$&\cr
\tablerule
&$3.4$&&$+8.381.10^{-3}$&&$+8.378.10^{-3}$&&$-5.947.10^2$&&$-4.192.10^{18}$&\cr
\tablerule
&$3.6$&&$+1.143.10^{-2}$&&$+1.142.10^{-2}$&&$0$&&$-3.812.10^{18}$&\cr\tablerule
&$3.8$&&$+1.381.10^{-2}$&&$+1.381.10^{-2}$&&$+1.867.10^2$&&$-3.454.10^{18}$&\cr
\tablerule
&$4.0$&&$+1.563.10^{-2}$&&$+1.563.10^{-2}$&&$+2.649.10^2$&&$-3.126.10^{18}$&\cr
height2pt&\omit&&\omit&&\omit&&\omit&&\omit&\cr}
\hrule}

\sect{{\bf 3. THE POLAR PERTURBATIONS, AND THE
MAGNETICALLY--CHARGED BLACK HOLE}}

\noindent{The analysis of the polar perturbations of the
electrically--charged string--theoretic black hole, as well as the
analysis of both the polar and the axial perturbations of the
associated magnetically--charged black hole, poses much more formidable
problems than we have dealt with until now. In
particular, for the magnetically--charged black hole one faces a task
of considerable grimness. The polar perturbations of both types of
black hole are difficult to analyze, in part due to the complicated
nature of the
partial differential equation which arises upon perturbing the dilaton
equation of motion (eq.(22), in the case of
the electrically--charged black hole,
and the analogous equation which we shall display presently
(eq.(92)) in the case
of the magnetically--charged black hole.). Furthermore there are
{\it more} polar potentials to solve for than there are axial
potentials, since, as we have pointed out, there are only
gravitational and electromagnetic axial perturbations while there
are gravitational, electromagnetic and dilatonic polar perturbations.
Even the
analysis of the equations for the
axial{\footnote {We will see that for the magnetic black hole the
notions of `axial' and `polar' perturbations take on different meanings
from the electric case: they are not exchanged precisely
{\it mutatis mutandis},
however, but rather interwoven in a special way
due to the symplectic transformation of the electric
field, as we shall
see.}}
perturbations of the magnetic
black hole is rendered difficult due to the characteristic angular dependence
of the magnetic field. In all cases the problem of separating the variables
in the partial differential equations, not to speak of decoupling the
equations, is a particularly troublesome issue.}

\noindent{As we will defer to a subsequent article
a comprehensive analysis of these perturbation problems of
four--dimensional string--theoretic black holes [30], here we will only
outline the details of the problem. We will set forth the relevant
perturbation equations, and, in the case of the axial perturbations of
the magnetically--charged black hole, indicate how one is led to analyze
a pair of coupled partial integrodifferential equations - which
can in fact be
subjected, and made to yield, to further analysis.}

\noindent{We take now as our starting point the original equations
of motion (eq.'s (2) through (4)) derived from the action in eq.(1)
(supplemented with eq.(5)),
for which we have a magnetically--charged black hole solution
characterized by the electromagnetic field--strength tensor
$F_{\mu\nu}$ given in
eq.(8). The metric components and the dilaton field will be taken as
in eq.'s (7) and (9) through (10), the relative plus sign in eq.(3)
indicating that we are working with the proper magnetic solution. We
may note
first that the Maxwell equations appropriate to this case will be
different from those listed in eq.'s (20) and (21), as the dilaton
field no longer accompanies the dual field strength in the basic
equations (eq.'s (2) and (5)) of the problem. Taking account of this
fact yields the following expressions for the Maxwell equations:}

$$\left(e^{f_1+f_2}\d F_{12}\right)_{,3}+\left(e^{f_1+f_3}F_{
31}\right)_{,2}=0 ~,\eqno(74a)$$

$$\left(e^{f_1+f_0}\d F_{01}\right)_{,2}+\left(e^{f_1+f_2}\d F_{
12}\right)_{,0}=0 ~,\eqno(74b)$$

$$\left(e^{f_1+f_0}\d F_{01}\right)_{,3}+\left(e^{f_1+f_3}F_{
13}\right)_{,0}=0 ~,\eqno(74c)$$

$$\eqalignno{\left(e^{f_2+
f_3}\d F_{01}\right)_{,0}&+\left(e^{f_0+f_3}\d F_{
12}\right)_{,2}+\left(e^{f_0+f_2}F_{13}\right)_{,3}\cr
&=e^{f_1+f_3}
\d F_{02}\Xi_{02}+e^{f_1+f_2}\d F_{03}\Xi_{03}-e^{f_1+f_0}
\d F_{23}\Xi_{23}\cr &+2\left(e^{f_2+f_3}\d F_{01}\Phi_{,0}
+e^{f_0+f_3}\d F_{12}\Phi_{,2}+e^{f_0+f_2}F_{13}\Phi_{
,3}\right) ~,&(74d)\cr}$$

\noindent{and}

$$\left(e^{f_1+f_3}\d F_{02}\right)_{,2}+\left(e^{f_1+f_2}\d F_{
03}\right)_{,3}=+2\left(e^{f_1+f_3}\d F_{02}\Phi_{,2}+e^{f_1+f_2}
\d F_{03}\Phi_{,3}\right) ~,\eqno(75a)$$

$$-\left(e^{f_1+f_0}\d F_{23}\right)_{,2}+\left(e^{f_1+f_2}\d F_{
03}\right)_{,0}=-2\left(e^{f_1+f_0}\d F_{23}\Phi_{,2}-e^{f_1+f_2}
\d F_{03}\Phi_{,0}\right) ~,\eqno(75b)$$

$$\left(e^{f_1+f_0}\d F_{23}\right)_{,3}+\left(e^{f_1+f_3}\d F_{
02}\right)_{,0}=+2\left(e^{f_1+f_0}\d F_{23}\Phi_{,3}
+e^{f_1+f_3}\d F_{
02}\Phi_{,0}\right) ~,\eqno(75c)$$

$$\eqalignno{\left(e^{f_0+f_2}\d F_{02}\right)_{,3}&-\left(e^{f_0+
f_3}\d F_{03}\right)_{,2}+\left(e^{f_2+f_3}\d F_{23}
\right)_{,0}\cr
&=e^{f_1+f_0}\d F_{01}\Xi_{23}+e^{f_1+f_2}\d F_{12}\Xi_{03}-
e^{f_1+f_3}F_{13}\Xi_{02} ~.&(75d)\cr}$$

\noindent{Bearing in mind that now $F_{13}=-Q\sin\theta$ is the
only independent non--vanishing component of the Maxwell tensor, it
is straightforward to linearize the preceeding equations, and we
readily obtain:}

$$\left(e^{f_0+f_3}\d F_{01}\right)_{,r}+e^{-f_0+f_3}\d F_{12,0}
=0 ~,\eqno(74b')$$

$$e^{f_0+f_3}\left(\d F_{01}\sin\theta\right)_{,\theta}+e^{2f_3}
\left[\d F_{13,0}-Q\left(\d f_1+\d f_3\right)_{,0}\sin\theta\right]
\sin\theta=0 ~,\eqno(74c')$$

$$\eqalignno{e^{-f_0+f_3}&\d F_{01,0}+\left(e^{f_0+f_3}\d F_{12}\right)_{
,r}+\left[\d F_{13}-Q\left(\d f_0+\d f_2\right)\sin\theta\right]_{
,\theta}\cr &=+2\left(e^{f_0+f_3}\d F_{12}\Phi_{,r}+F_{13}\d\Phi_{
,\theta}\right) ~,&(74d')\cr}$$

\noindent{and}

$$\left(e^{2f_3}\d F_{02}\right)_{,r}+e^{-f_0+f_3}\left(\d F_{03}
\sin\theta\right)_{,\theta}\csc\theta=+2e^{2f_3}\d F_{02}\Phi_{,r}
{}~,\eqno(75a')$$

$$-\left(e^{f_0+f_3}\d F_{23}\right)_{,r}+e^{-f_0+f_3}\d F_{03,0}=
-2e^{f_0+f_3}\d F_{23}\Phi_{,r} ~,\eqno(75b')$$

$$e^{f_0+f_3}\left(\d F_{23}\sin\theta\right)_{,\theta}+e^{2f_3}
\d F_{02,0}\sin\theta=0 ~,\eqno(75c')$$

$$\d F_{02,\theta}-\left(e^{f_0+f_3}\d F_{03}\right)_{,r}+e^{
-f_0+f_3}\d F_{23,0}=-e^{2f_3}F_{13}\Xi_{02}\sin\theta ~,\eqno(75d')$$

\noindent{where now eq.(74$a$) serves as the integrability
condition for eq.'s (74$c$) and (74$c$), and we
have therefore not linearized
it. Inspection of eq.'s (74$'$) and (75$'$) reveals the `loss of identity'
on the part of the polar and axial perturbations, respectively,
referred to in the footnote above. We see that when the Maxwell equations
for the perturbed field components are grouped in the same way as
was done for the electric case, so that those equations
which are invariant to the substitution
$\varphi\rightarrow -\varphi$ are separated from those which change
sign, the anticipated dynamical evolution is no
longer determined by the {\it linearized} version of
each set. In eq.'s (74$'$) we find terms containing
true polar perturbations (such as $\d f_1$)
as well as terms containing true axial perturbations (such as $\d F_{01}$),
and in the case of eq.'s (75$'$) we again encounter both
polar perturbations
(such as $\d F_{02}$) and axial perturbations (such as $\Xi_{02}$).
We see that
the special character of the ($\Phi$--dependent) duality transformation
makes the
change from the electric equations to the magnetic equations more
complicated than the simple `switch' polar$\rightleftharpoons$axial
would entail.}

\noindent{In order to obtain the
magnetic Einstein equations for the perturbed fields we will
need, as before, the appropriate components of the perturbed
stress--energy tensor, which is given now by ({\it cf.} eq.(26))}

$$\eqalignno{\d T_{ab}^{({\rm tot})}=&2\left(\d\Phi_{,a}\Phi_{,b}+
\Phi_{,a}\d\Phi_{,b}\right)\cr &+2e^{-2\Phi}\left[\eta^{cd}\left(
\d F_{ac}F_{bd}+F_{ac}\d F_{bd}\right)-\eta_{ab}F_{13}\d F_{13}
-T_{ab}^{({\rm EM})}\d\Phi\right] ~,&(76)\cr}$$

\noindent{from which we may obtain the relevant quantities
recorded below:}

$$\d T_{13}^{({\rm tot})}=0 ~,\eqno(77)$$

$$\d T_{12}^{({\rm tot})}=-2e^{-2\Phi}Q\d F_{23}\sin\theta ~,\eqno(78)$$

$$\d T_{02}^{({\rm tot})}=+2\Phi_{,r}\d\Phi_{,0} ~,\eqno(79)$$

$$\d T_{03}^{({\rm tot})}=+2e^{-2\Phi}Q\d F_{01}\sin\theta ~,\eqno(80)$$

$$\d T_{23}^{({\rm tot})}=+2\Phi_{,r}\d\Phi_{,\theta}+2e^{-2\Phi}
Q\d F_{21}\sin\theta ~,\eqno(81)$$

$$\d T_{11}^{({\rm tot})}=-2e^{-2\Phi}\left(Q\d F_{13}\sin\theta+
Q^2\d\Phi\sin^2\theta\right) ~,\eqno(82)$$

$$\d T_{22}^{({\rm tot})}=4\Phi_{,r}\d\Phi_{,r}+2e^{-2\Phi}\left(
Q\d F_{13}\sin\theta+Q^2\d\Phi\sin^2\theta\right) ~,\eqno(83)$$

\noindent{where we have made use of the appropriate electromagnetic
stress--energy tensor ({\it cf.} eq.(27)):}

$$T_{ab}^{({\rm EM})}=2F_{ac}F_{b}^{~c}-\eta_{ab}Q^2\sin^2\theta
{}~.\eqno(84)$$

\noindent{With the preceeding expressions we derive the various
relevant components of the perturbed Einstein equations:}

$$\displaylines{\d R_{12}=\d T_{12}\Rightarrow\hfill\cr}$$
$$\left(e^{2f_0+2f_3}\Xi_{23}\sin^3\theta\right)_{,\theta}-
e^{4f_3}\Xi_{20,0}\sin^3\theta=-4Qe^{-2\Phi}e^{f_0+3f_3}
\d F_{23}\sin^3\theta ~,\eqno(85)$$

$$\displaylines{\d R_{13}=\d T_{13}\Rightarrow\hfill\cr}$$
$$\left(e^{2f_0+2f_3}\Xi_{23}\right)_{,r}-e^{-2f_0+2f_3}\Xi_{03,0}
=0 ~,\eqno(86)$$

$$\displaylines{\d R_{22}=\d T_{22}\Rightarrow\hfill\cr}$$
$$\eqalignno{&e^{2f_0}\left[2{f_3}_{,r}\d {f_0}_{,r}+\left({f_0}_{,r}+
{f_3}_{,r}\right)\left(\d f_1+\d f_3\right)_{,r}-2\left(2{f_0}_{,r}
{f_3}_{,r}+{f_3}_{,r}^{~2}\right)\d f_2\right]\cr &+e^{-2f_3}\left[
\left(\d f_1+\d f_0\right)_{,\theta ,\theta}+\left(\d f_0-\d f_3+
2\d f_1\right)_{,\theta}\cot\theta+2\d f_3\right]\cr &-e{-2f_0}
\left(\d f_1+\d f_3\right)_{,0,0}\cr &=4\Phi_{,r}\d\Phi_{,r}+
2e^{-2\Phi}\left(Q\d F_{13}\sin\theta+Q^2\d\Phi\sin^2\theta
\right) ~,&(87)\cr}$$

$$\displaylines{\d R_{11}=\d T_{11}\Rightarrow\hfill\cr}$$
$$\eqalignno{&e^{2f_0}\{\d {f_1}_{,r,r}+{f_3}_{,r}
(\d f_1+\d f_0+\d f_3-\d f_2)_{,r}\cr &+2\d {f_1}_{,r}
({f_3}_{,r}+{f_1}_{,r})-2[{f_3}_{,r,r}+2{f_3}_{,r}
({f_3}_{,r}+{f_0}_{,r})]\d f_2\}\cr
&+e^{-2f_3}[\d {f_1}_{,\theta ,\theta}+(2\d f_1+\d f_0
+\d f_2-\d f_3)_{,\theta}\cot\theta +2\d f_3]\cr
&-e^{-2f_0}\d {f_1}_{,0,0}\cr &=2e^{-2\Phi}(Q
\d F_{13}\sin\theta+Q^2\d\Phi\sin^2\theta) ~,&(88)\cr}$$

$$\displaylines{\d R_{23}=\d T_{23}\Rightarrow\hfill\cr}$$
$$\eqalignno{\left({f_0}_{,r}+{f_3}_{,r}\right)\d {f_2}_{,\theta}
&-\left(\d f_1+\d f_0\right)_{,r,\theta}-\left({f_0}_{,r}-{f_3}_{,r}
\right)\d {f_0}_{,\theta}-\left(\d f_1-\d f_3\right)_{,r}\cot\theta\cr
&=2e^{-f_0+f_3}\left(\Phi_{,r}\d\Phi_{,\theta}+e^{-2\Phi}Q
\d F_{21}\sin\theta\right) ~,&(89)\cr}$$

$$\displaylines{\d R_{03}=\d T_{03}\Rightarrow\hfill\cr}$$
$$\left(\d f_1+\d f_2\right)_{,\theta,0}+\left(\d f_1-\d f_3\right)_{,0}
\cot\theta=-2e^{-2\Phi}e^{f_0+f_3}Q\d F_{01}\sin\theta ~,\eqno(90)$$

$$\displaylines{\d R_{02}=\d T_{02}\Rightarrow\hfill\cr}$$
$$\left[\left(\d f_1+\d f_3\right)_{,r}+\left(\d f_1+\d f_3\right)\left(
{f_3}_{,r}-{f_0}_{,r}\right)-2{f_3}_{,r}\d f_2\right]_{,0}
=-2\Phi_{,r}\d\Phi_{,0} ~.\eqno(91)$$

\noindent{We proceed to
obtain the equation which is produced upon perturbing
the dilaton equation of motion (eq.(3)), for which we must take into
account
now the relative sign difference between the two terms which distinguishes
this equation from the version appropriate to the electric black hole
(eq.(3$''$)). However, the derivation is clear if we
recall that the tilde--duality transformation
requires changing the sign of the dilaton field, and continue to bear
in mind that it is $F_{13}$ now, and not ${\cal F}_{02}$,
which is nonvanishing. Carrying through the analysis we obtain:}

$$\eqalignno{&e^{-2f_0}\delta\Phi_{,0,0}-e^{-2f_2}\left[\delta\Phi_{,r,r}
+\left(f_1+f_0-f_2+f_3\right)_{,r}\delta\Phi_{,r}\right]\cr &-e^{-2f_3}
\left[\delta\Phi_{,\theta,\theta}+\left(f_1+f_0+f_2
-f_3\right)_{,\theta}\delta\Phi_{,\theta}
\right]-2e^{-2\Phi}e^{-2f_0-2f_2}F_{13}^{~2}\delta\Phi\cr
&+e^{-2f_2}\left\{\left(\delta f_1+\delta f_0-\delta f_2+\delta f_3
\right)_{,r}\Phi_{,r}-2\delta f_2\left[\Phi_{,r,r}+\left(f_1+f_0-f_2+
f_3\right)_{,r}\Phi_{,r}\right]\right\}\cr &+2e^{-2\Phi}e^{-2f_0-2f_2}
\left[F_{13}\delta F_{13}-\left(\delta f_0+\delta f_2\right)
F_{13}^{~2}\right]=0 ~.&(92)\cr}$$

\noindent{With the foregoing equation for the dilaton perturbation
we again observe that only ``polar" perturbations appear, and once
again we will find that the equations
for the ``axial"
perturbations (see below) do not determine the evolution
of the perturbation in the dilaton field.}

\noindent{We are now in a position to
write down the ``axial" and ``polar" perturbation
equations for the magnetic black hole analogous to those of the
electric black hole. Considering as before the ``axial" perturbations,
we can carry through the analysis leading up to the equations
analogous to eq.(44) and eq.(45) for the black electric pole. We
derive in this way:}

$$\eqalignno{e^{-f_0+f_3}\g_{,0,0}&-
{\partial\over \partial r}\left[e^{2f_0}
{\partial\over \partial r}\left( e^{f_0+f_3}\g\right)\right]
+2{\partial\over \partial r}\left( e^{3f_0+f_3}\g\Phi_{,r}\right)\cr
&-e^{f_0-f_3}
{\partial\over \partial\theta}\left( {1\over \sin\theta}{\partial\g\over
\partial\theta}\right)
\sin\theta=-e^{2f_3}F_{13}\Xi_{02,0}\sin^2\theta ~,&(93)\cr}$$

\noindent{by appropriately combining eq.'s (75$b'$) through (75$d'$)
to eliminate $\d F_{02}$ and $\d F_{03}$ in
favor of $\d F_{23}$ and
we have set $\d F_{23}\sin\theta\equiv \g$;
and}

$$\eqalignno{e^{4f_3}{\partial\over \partial r}\left(e^{-2f_3+2f_0}
{\partial\a\over \partial r}\right)+\sin^3\theta{\partial\over \partial
\theta}\left({1\over \sin^3\theta}{\partial\a\over \partial\theta}\right)
&-\a_{,0,0}e^{2f_3-2f_0}\cr &=2e^{f_0+3f_3}\sin^3\theta{\partial\over
\partial\theta}\left({\delta T_{12}\over \sin\theta}\right) ~,&(94)\cr}$$

\noindent{by combining the $(1,2)$-- and $(1,3)$--components of the
perturbed Einstein equations, and where,
as before, we have defined $e^{2f_3+2f_0}\Xi_{23}
\sin^3\theta\equiv \a$, and $\d T_{12}$ is as given in eq.(78).}

\noindent{We now note that, since eq.(85) implies}

$$\Xi_{02,0}=2e^{f_0-f_3}\d T_{12}\csc\theta-e^{-4f_3}{\partial\a\over
\partial\theta}\csc^3\theta ~,\eqno(95)$$

\noindent{the use of eq.(78) yields from eq.'s (93) and (94), respectively}

$$\eqalignno{2{\partial\over \partial r}\left(e^{3f_0
+f_3}\g\Phi_{,r}\right)&-
{\partial\over \partial r}\left[e^{2f_0}{\partial\over \partial r}
\left(e^{f_0+f_3}\g\right)\right]-e^{-f_0+f_3}\omega^2\g-e^{f_0-f_3}
{\partial\over \partial\theta}\left({1\over \sin\theta}{\partial\g\over
\partial\theta}\right)\sin\theta\cr &=-4Q^2e^{-2\Phi}e^{f_0+f_3}\g\sin^2
\theta-Qe^{-2f_3}{\partial\a\over \partial\theta} ~,&(96)\cr}$$

\noindent{and}

$$\eqalignno{e^{4f_3}{\partial\over \partial r}\left(e^{2f_0-2f_3}
{\partial\a\over \partial r}\right)+&{\partial\over \partial\theta}
\left({1\over \sin^3\theta}{\partial\a\over \partial\theta}\right)\sin^3
\theta+\omega^2e^{-2f_0+2f_3}\a\cr &=-4Qe^{-2\Phi}e^{f_0+3f_3}{\partial\over
\partial\theta}\left({\g\over \sin\theta}\right)\sin^3\theta ~,&(97)\cr}$$

\noindent{Unfortunately, these equations
do not appear to admit the consistent, {\it simultaneous}
separation of variables which
characterizes the corresponding equations for the electric black hole.
Different choices of definition for $\g$ do not ameliorate the
situation,{\footnote {One is in any
case constrained to define $\g$ as {\it some}
linear function of $\d F_{23}$ since this quantity is what necessarily
appears within the expression $\Xi_{02}$ on the right hand side of the
linearized Maxwell equation (75$d'$).}} which is essentially a consequence
of the appearance in the $(1,2)$--component of the perturbed stress--tensor
of {\it both} a radial and a manifest angular
dependence, in contrast to the situation
for the electrically--charged black hole ({\it cf.} eq.'s (78) and (29)).}

\noindent{Therefore, we proceed in the following manner. We may consider
the general coupled pair of partial differential equations given by}

$${\hat {\cal H}}_1(x,y)\psi_1(x,y)={\hat f}_1(x,y)\psi_2(x,y) ~,\eqno(98)$$

\noindent{and}

$${\hat {\cal H}}_2(x,y)\psi_2(x,y)={\hat f}_2(x,y)\psi_1(x,y) ~,\eqno(99)$$

\noindent{where ${\hat {\cal H}}_1$, ${\hat {\cal H}}_2$, ${\hat f}_1$
and ${\hat f}_2$ are given differential operators, and we
introduce Green functions $G_1$ and $G_2$
for ${\hat {\cal H}}_1$ and ${\hat {\cal H}}_2$ by writing}

$${\hat {\cal H}}_1(x,y)G_1(x,x';y,y')=\d (x,x')\d (y,y') ~,\eqno(100)$$

\noindent{and}

$${\hat {\cal H}}_2(x,y)G_2(x,x';y,y')=\d (x,x')\d (y,y') ~.\eqno(101)$$

\noindent{Since the solution for, say, $\psi_1$ is given by}

$$\psi_1(x,y)=\psi_1^{(0)}(x,y)+\int\int dx' dy' G_1(x,x';y,y'){\hat f}_1
(x',y')\psi_2(x',y') ~,\eqno(102)$$

\noindent{where $\psi_1^{(0)}$ is the homogeneous solution of eq.(98),
we can recast eq.(99) as the following linear partial
integrodifferential equation}

$$\eqalignno{{\hat {\cal H}}_2\psi_2(x,y)=&{\hat f}_2(x,y)\psi_1^{(0)}
(x,y)\cr &+{\hat f}_2(x,y)\int\int dx' dy' G_1(x,x';y,y'){\hat f}_1(
x',y')\psi_2(x',y') ~,&(103)\cr}$$

\noindent{and we obtain a similar equation for $\psi_1$ by
interchanging the indices $1$ and $2$. These linear
integrodifferential equations are amenable to numerical analysis,
but for analytical consideration
it is more useful to rewrite them by substituting
back into eq.'s (98) and (99), after which we obtain:}

$$\psi_1(x,y)=h_1(x,y)+\int\int dx' dy' {\cal K}_1(x,x';y,y')
\psi_1(x',y') ~,\eqno(104)$$

\noindent{and}

$$\psi_2(x,y)=h_2(x,y)+\int\int dx' dy' {\cal K}_2(x,x';y,y')
\psi_2(x',y') ~\eqno(105)$$

\noindent{where}

$$h_1(x,y)\equiv \psi_1^{(0)}(x,y)+\int\int dx' dy' G_1(x,x';y,y')
{\hat f}_1(x,x')\psi_2^{(0)}(x',y') ~,\eqno(106)$$

\noindent{and}

$${\cal K}_1(x,x'';y,y'')\equiv \int\int dx' dy' G_1(x,x';y,y'){\hat
f}_1(x',y')G_2(x',x'';y',y''){\hat f}_2(x'',y'') ~,\eqno(107)$$

\noindent{and similar definitions obtain for
$h_2$ and ${\cal K}_2$
with the
interchange of the indices $1$ and $2$. The linear integral
equations (104) and (105) so obtained are Fredholm equations of
the second kind,
for which we may make use of the large body of established
theorems regarding the existence and asymptotic behavior of
their solutions [31,32,33]. For the coupled partial differential equations
characterizing the ``axial" perturbations of the magnetic black
hole (eq.'s (96) and (97)), comparison with eq.'s (98) and (99) reveals that
we should make the identifications}

$$\eqalignno{{\hat {\cal H}}_1(r,\theta)\equiv &{\partial^2\over \partial
r^2}-2\left(\Phi-2f_0-f_3\right)_{,r}{\partial\over \partial r}\cr
&-2\left[\Phi_{,r,r}+\left(3f_0+f_3\right)_{,r}\Phi_{,r}\right]+
\left(f_0+f_3\right)_{,r,r}+\left(3f_0+f_3\right)_{,r}\left(f_0+f_3\right)_{
,r}\cr &-e^{-4f_0}\omega^2+4Q^2e^{-2\Phi}e^{-2f_0}\sin^2\theta\cr
&+e^{-2f_0-2f_3}\left({\partial^2\over \partial\theta^2}-\cot\theta{\partial
\over \partial\theta}\right) ~,&(108)\cr}$$

$${\hat f}_1(r,\theta)\equiv Qe^{-3f_0-3f_3}{\partial\over
\partial\theta} ~,\eqno(109)$$

$$\eqalignno{{\hat {\cal H}}_2(r,\theta)\equiv &{\partial^2\over
\partial r^2}+2\left(f_0-f_3\right)_{,r}{\partial\over
\partial r}+e^{-4f_0}\omega^2\cr
&+e^{-2f_0-2f_3}\left({\partial^2\over \partial\theta^2}-3\cot
\theta{\partial\over \partial\theta}\right) ~,&(110)\cr}$$

\noindent{and}

$${\hat f}_2(r,\theta)\equiv -4Qe^{-2\Phi}e^{-f_0+f_3}\sin^2\theta\left(
{\partial\over \partial\theta}-\cot\theta\right) ~.\eqno(111)$$

\noindent{The kernels ${\cal K}_1$ and ${\cal K}_2$ are completely
specified in terms of the preceeding expressions, and the inhomogeneity
terms $h_1$ and $h_2$ of the Fredholm equations can be obtained once the
homogeneous solutions ($\psi_1^{(0)}$ and $\psi_2^{(0)}$)
of the partial differential equations (eq.'s (96) and (97))
are computed, which can be done
in a straightforward manner using the method of Frobenius.{\footnote
{Of course, the separation of variables proceeds without difficulty in
the case of the homogeneous forms of the equations.}}
The solutions
to the integral equations can then be studied by repeated iteration. [33]}

\sect{{\bf 4. TWO--DIMENSIONAL BLACK HOLES}}

\noindent{We will consider the analysis of the perturbations of
two--dimensional string--theoretic black holes in a forthcoming
article, and content ourselves here with providing a listing of
the relevant equations for a recently discovered charged configuration
[8]. We first note that, in
the case of two--dimensional black holes the
most sufficiently general form for the perturbed metric associated
with a given initial configuration can be represented as a
diagonal matrix. This is a consequence of the fact that, for either
Lorentzian or Euclidean signature, it is always possible to bring
a two--dimensional metric into diagonal form through a transformation
of the coordinates. This means that, in
the analysis of the first--order perturbations of two--dimensional
black holes, we will not encounter the analogues of ``axial" perturbations;
only two--dimensional ``polar" perturbations will arise, for which the
sign of the metric remains unchanged as the signs of the perturbations are
reversed. Thus the metric corresponding to the squared line element}

$$ds^2=-e^{2f_0}dt^2+e^{2f_1}dr^2 ~,\eqno(112)$$

\noindent{will experience perturbations in the form}

$$\pmatrix{-e^{2f_0}&0\cr 0&e^{2f_1}\cr}~~\rightarrow ~~\pmatrix{
-e^{2f_0+2\d f_0}&0\cr 0&e^{2f_1+2\d f_1}\cr} ~.\eqno(113)$$

\noindent{It is clear that a great simplification appears which
distinguishes the analysis of two--dimensional black holes from
the analysis appropriate to higher--dimensional black holes. The
various equations for the different perturbations
are automatically
separated in the coordinates, if we continue to assume that all
field perturbations carry a time--dependence $\propto e^{i\omega t}$.
Nevertheless, the fact that the perturbations are ``polar" brings
with it the difficulties attendant to decoupling the
separated differential equations.}

\noindent{We shall give here a brief survey of the initial steps
required for the perturbation analysis of the charged two--dimensional
black hole [8] described by the metric tensor characterized by the
metric components:}

$$g_{00}=-e^{2f_0}=-2k\tanh^2r ~,\eqno(114)$$

$$g_{11}=e^{2f_1}=2k ~,\eqno(115)$$

\noindent{where $k$ is the level of the Wess--Zumino action, and
for which the dilaton field is given by}

$$\Phi=\log\cosh^2r+a ~,\eqno(116)$$

\noindent{where $a$ is a constant. The two--form field strength
of the gauge field which
characterizes the black hole solution has in two dimensions, of
course, only one
independent non--vanishing component, given by}

$$F_{01}=-2Q\sinh r~{\rm sech}^3~r ~,\eqno(117)$$

\noindent{and the `electric/magnetic'
charge of the gauge field is $Q=2\e k^{-1}e^a$,
where $\e$ is the coupling between the worldsheet gauge field and
a free boson field.}

\noindent{We proceed to consider the perturbation of the electromagnetic
equations given by}

$$\nabla_\mu\left(e^{\Phi}F^{\mu\nu}\right)=0 .~\eqno(118)$$

\noindent{In two dimensions this yields only one independent
equation, from which we derive}

$$\left(\Phi_{,r}-{f_0}_{,r}\right)\d F_{01}+F_{01}\d\Phi_{,r}
-F_{01}\left(\d f_0+\d f_1\right)_{,r}+\d F_{01,r}=0 ,~\eqno(119)$$

\noindent{which can be rewritten as}

$$\left(2\sinh^2r-1\right)\d F_{01}+\d F_{01,r}\sinh r\cosh r=
2Q\left(\d\Phi-\d f_0-\d f_1\right)_{,r}\tanh^2r .~\eqno(120)$$

\noindent{Expressing the components of the Ricci tensor in terms
of arbitrary static
metric functions $f_0$ and $f_1$ we find (recall that
we now define $R_{\mu\nu}=R^\rho_{~\mu\nu\rho}$)}

$$R_{00}=-e^{2f_0-2f_1}\left({f_0}_{,r,r}+{f_0}_{,r}^{~2}-{f_1}_{
,r}{f_0}_{,r}\right) ,~\eqno(121)$$

\noindent{and}

$$R_{11}={f_0}_{,r,r}+{f_0}_{,r}^{~2}-{f_1}_{,r}{f_0}_{,r} .~\eqno(122)$$

\noindent{The corresponding components of the stress-energy tensor of
the problem may be readily derived as well.
Effecting the linearization in the perturbations $\d f_0$
and $\d f_1$ and setting the relevant quantities equal to each other
we obtain the perturbed Einstein equations}

$$\eqalignno{-e^{2f_0-2f_1}&\left[\d {f_0}_{,r,r}+\left(2{f_0}_{,r}
-{f_1}_{,r}\right)\d {f_0}_{,r}-{f_0}_{,r}\d {f_1}_{,r}+2\left(
\d f_0-\d f_1\right)\left({f_0}_{,r,r}+{f_0}_{,r}^{~2}-{f_1}_{,r}
{f_0}_{,r}\right)\right]\cr &=4Q\d F_{01}\sinh r~{\rm sech}^3 r ~,&(123)\cr}$$

\noindent{and}

$${f_0}_{,r,r}+\left(2{f_0}_{,r}-{f_1}_{,r}\right)\d {f_0}_{,r}
-{f_0}_{,r}\d {f_1}_{,r}=\Phi_{,r}\d\Phi_{,r}-4Q\d F_{01}\sinh r
{}~{\rm sech}^3 r ~.\eqno(124)$$

\noindent{We may combine these two equations, which, with the
field values given in eq.'s (114) through (117), reduce to}

$$\eqalignno{\d {f_0}_{,r,r}+2\left(\d f_0-\d f_1\right)_{,r}{\rm
sech}~r
{}~{\rm csch}~r-4\left(\d f_0-\d f_1\right)&\tanh^2r\cr
&=2\d\Phi_{,r}\sinh r\cosh r ~.&(125)\cr}$$

\noindent{Similarly, the somewhat more complicated perturbed form of the
dilaton equation can be verified to be given by}

$$\eqalignno{\omega^2&\d\Phi\coth^2r+\d\Phi_{,r,r}+\d\Phi_{,r}{\rm sech}~r
{}~{\rm csch}~r+2Q\e\d\Phi{\rm sech}^2r\cr =&2Q\left[\d F_{01}\coth r+2Q\left(
\d f_0+\d f_1\right){\rm sech}^2r\right]-2\left(\d f_0-\d f_1\right)_{,r}
\tanh r\cr &+8\d f_1{\rm sech}^2r ~,&(126)\cr}$$

\noindent{where we have assumed the usual time--dependence. We may
note that, if we make the field redefinition $\d\Phi=\left(\tanh r
\right)^{-1/2}P(r)$, we can rewrite the perturbed
dilaton equation without the first--order derivative term as}

$$\left({d^2\over dr^2}+\omega^2\coth^2r\right)P(r)=f(r)P(r)+
\left(\tanh r\right)^{1/2}K(r) ~,\eqno(127)$$

\noindent{where}

$$f(r)\equiv {\rm sech}^2r-{1\over 4}{\rm sech}^2r~{\rm csch}^2r+
{\rm csch}^2r+2Q\e~{\rm sech}^2r ~,\eqno(128)$$

\noindent{and}

$$\eqalignno{K(r)\equiv &2Q\left[\d F_{01}\coth r+2\e\left(\d f_0+\d
f_1\right){\rm sech}^2r\right]-2\left(\d f_0-\d f_1\right)_{,r}\tanh r\cr
&+8\d f_1{\rm sech}^2r ~.&(129)\cr}$$

\noindent{A similar analysis can be carried out for the tachyonic
black hole which was
recently constructed in [7]. Work is in progress on these
problems [34].}

\newpage

\sect{{\bf 5. CONCLUDING REMARKS}}

\noindent{As we have observed, the dilaton field, and
more importantly the
perturbations in the dilaton field, appear throughout the
analysis of the string--theoretic
black hole in a very non--symmetrical fashion in the various
relevant equations. The significance of this can be best
appreciated by considering the situation which prevails for
the static black holes of general relativity. It is a noteworthy
fact that for these black hole configurations the potentials for
the axial and the polar perturbations are interrelated in a very
special and non--trivial manner. The important point is that the
scattering behavior of the
black holes is highly constrained, with the reflection and
transmission coefficients associated with the axial and polar
perturbations being precisely equal to each other [27,28]. Although this
fact is not the same as the fact of the stability of these black
holes, there is a relation between the two. Although we will
carefully consider the corresponding issues of scattering and
stability for string--theoretic black holes in a separate
paper [30], we may observe that one should expect entirely new
identities relating the scattering coefficients, and the bounds
constraining the energy integrals which govern stability, in
string theory.}

\noindent{(We may note in passing that, should incontrovertible
evidence in favor of an astrophysical black hole be obtained,
it {\it may} be possible to use some analogue of
our expressions for the perturbation potentials{\footnote {This because
it should be rather unlikely to encounter an astrophysical black hole with
a net charge.}}
to set approximate constraints
on the asymptotic value of the dilaton field.
For, if one seriously
advocates string theory as the correct underlying
theoretical description, bearing
in mind that black holes are (essentially uniquely
among macroscopic objects) more or less exactly described in terms of
a fundamental theory, a definitive observation of a black hole with
certain measured characteristics may serve to exclude certain values
for the asymptotic dilaton field strength if these correspond to
theoretically unstable configurations. Of course, this is highly
speculative, and it should be further pointed out that such ``constraints"
would be approximate, owing as much to the extreme difficulty of securing
an {\it exact} description of string theory as to the typical vagaries
of astronomical measurement.)}

\noindent{It is important to carefully analyze the perturbations
which characterize the other string--theoretic black holes which have
been constructed. For instance, the four--dimensional dual, dilaton/axion
black hole [9] which is an electromagnetic dyon should be subjected to
perturbation analysis. One might expect that the presence of the axion
together with the dilaton may restore a balance between the axial
and polar perturbations. However, it is important to take account
of the consequences of the higher--order correction
terms which have been ignored in this paper, in the case of both
two--dimensional and four--dimensional black holes. The inclusion
of such terms is likely to have significant implications for the behavior
of the perturbed configurations. For instance, we know that the
duality transformation satisfied by the dilaton field in the
one--loop approximation in the sigma--model is modified at two--loop
order [35], and the precise form of the corrections at
more than two loops
remains to be elucidated. We are daily learning to appreciate the
degree of the vastness which appears to be inherent to the study
of string theory: as we have shown, it is important to carry
out perturbation analyses of possible solutions to help further our
understanding.}

\noindent{{\it Acknowledgement:} The author wishes to thank Professor
Jim Gates for his kind, and crucial, encouragement.}

\newpage

\refs

\Item{[1]} E. Witten, {\it Phys. Rev.} {\bf D44} (1991) 314.
\Item{[2]} I. Bars, {\it String Propagation on Black Holes},
USC--91/HEP--B3, May 1991.
\Item{[3]} R. Dijkraaf, H. Verlinde and E. Verlinde, {\it String
Propagation in a Black Hole Geometry}, PUPT--1252; IASSNS--HEP--91/22,
May 1991.
\Item{[4]} C. Callan, R. Myers and M.J. Perry, {\it Nucl. Phys.}
{\bf B311} (1988) 673.
\Item{[5]} G.W. Gibbons and K. Maeda, {\it Nucl. Phys.} {\bf
B298} (1988) 741.
\Item{[6]} K. Lee and E. Weinberg, {\it Charge Black Holes with
Scalar Hair}, CU--TP--515, April 1991.
\Item{[7]} S.P. de Alwis and J. Lykken, {\it 2d Gravity and the
Black Hole Solution in 2d Critical String Theory}, COLO--HEP--258,
July 1991.
\Item{[8]} N. Ishibashi, M. Li and A.R. Steif, {\it Two Dimensional
Charged Black Holes in String Theory}, UCSBTH--91--28 (Revised
Version), July 1991.
\Item{[9]} A. Shapere, S. Trivedi and F. Wilczek, {\it Dual Dilaton
Dyons}, IASSNS--HEP--91/33, June 1991.
\Item{[10]} D. Garfinkle, G.T. Horowitz and A. Strominger,
{\it Phys. Rev.} {\bf D43} (1991) 3140.
\Item{[11]} B.A. Campbell, M.J. Duncan, N. Kaloper and K.A. Olive,
{\it Phys. Lett.} {\bf 251B} (1990) 34.
\Item{[12]} J.H. Horne and G.T. Horowitz, {\it Exact Black String
Solutions in Three Dimensions}, UCSBTH--91--39, July 1991.
\Item{[13]} S. Shapiro and S. Teukolsky, {\it Phys. Rev. Lett.} {\bf
66} (1991) 994.
\Item{[14]} K.A. Bronnikov and Y.N. Kireyev, {\it Phys. Lett.} {\bf
67A} (1978) 95.
\Item{[15]} T.J.M. Zouros and D.M. Eardley, {\it Ann. Phys} {\bf 118}
(1979) 139.
\Item{[16]} N. Straumann and Z. Zhou, {\it Phys. Lett.} {\bf 243B}
(1990) 33.
\Item{[17]} G.W. Gibbons, {\it Phys. Rev.} {\bf D15} (1977) 3530.
\Item{[18]} I. Semiz, {\it Class. Quantum Grav.} {\bf 7} (1990) 353.
\Item{[19]} D. Lohiya, {\it Phys. Rev.} {\bf D30} (1984) 1194.
\Item{[20]} W.A. Hiscock and L.D. Weems, {\it Phys. Rev.} {\bf D41}
(1990) 1142.
\Item{[21]} T. Regge and J.A. Wheeler, {\it Phys. Rev.} {\bf 108}
(1957) 1063.
\Item{[22]} F.J. Zerilli, {\it Phys. Rev.} {\bf D9} (1973) 860.
\Item{[23]} V. Moncrief, {\it Phys. Rev.} {\bf D9} (1974) 2707.
\Item{[24]}V. Moncrief, {\it Phys. Rev.} {\bf D10} (1974) 1057.
\Item{[25]} J. Bardeen, {\it Astrophys. J.} {\bf 161} (1970) 103.
\Item{[26]} S. Chandrasekhar and J.L. Friedman, {\it Astrophys. J.}
{\bf 175} (1972) 379.
\Item{[27]} S. Chandrasekhar, {\it Proc. Roy. Soc.} {\bf A365}
(1979) 453.
\Item{[28]} S. Chandrasekhar and B.C. Xanthopoulos,
{\it Proc. Roy. Soc.} {\bf A367} (1979) 1.
\Item{[29]} S. Wolfram, {\it Mathematica}, Addison--Wesley, 1988.
\Item{[30]} G. Gilbert, in preparation.
\Item{[31]} H. Hochstadt, {\it Integral Equations}, John Wiley \&
Sons, 1973.
\Item{[32]} M. Goldberger and K. Watson, {\it Collision Theory},
Krieger Publishing Company, 1975.
\Item{[33]} N. Muskhelishvili, {\it Singular Integral Equations},
P. Noordhoff, 1953.
\Item{[34]} G. Gilbert, in preparation.
\Item{[35]} A.A. Tseytlin, {\it On the Form of ``Black Hole" Solution
in $D=2$ String Theory}, JHU--TIPAC--91009 (Revised Version), July 1991.

\end{document}